\documentclass[aps,prd,twocolumn,showpacs,preprintnumbers,amsmath,amssymb, longbibliography]{revtex4-1}

\usepackage{bbm}
\usepackage{bm}
\usepackage{color}
\usepackage{dcolumn}
\usepackage{epstopdf}
\usepackage{graphicx}
\usepackage[colorlinks, citecolor=blue]{hyperref}
\usepackage{lipsum}
\usepackage{float}

\newcommand{\ket}[1]{|#1\rangle}

\newcommand{\fig}[1]{Fig.~\ref{#1}}

\newcommand{\eq}[1]{Eq.~(\ref{#1})}

\newcommand{\beq}{ \begin{equation} }
\newcommand{\eeq}{ \end{equation} }

\newcommand{\e}{\varepsilon}
\newcommand{\s}{\sigma}
\newcommand{\up}{\uparrow}
\newcommand{\down}{\downarrow}
\newcommand{\w}{\omega}
\newcommand{\Op}{\mathcal{O}}

\newcommand{\exch}{\Delta \varepsilon_{\rm exch}}
\newcommand{\exchn}{\exch^{\rm NRG}}

\begin{document}

\title{Quench dynamics of spin in quantum dots coupled to spin-polarized leads}

\author{Kacper Wrze\'sniewski}
\email{wrzesniewski@amu.edu.pl}

\author{Ireneusz Weymann}
\email{weymann@amu.edu.pl}
\affiliation{Faculty of Physics, Adam Mickiewicz University, Umultowska 85, 61-614 Pozna{\'n}, Poland}


\begin{abstract}
We investigate the quench dynamics of a quantum dot
strongly coupled to spin-polarized ferromagnetic leads.
The real-time evolution is calculated by means of the time-dependent density-matrix numerical
renormalization group method implemented within the matrix product states framework.
We examine the system's response to a quench in the spin-dependent coupling strength to
ferromagnetic leads as well as in the position of the dot's orbital level.
The spin dynamics is analyzed by calculating the time-dependent expectation values
of the quantum dot's magnetization and occupation.
Based on these, we determine the time-dependence of a ferromagnetic-contact-induced
exchange field and predict its nonmonotonic build-up.
In particular, two time scales are identified,
describing the development of the exchange field and the dot's magnetization sign change.
Finally, we study the effects of finite temperature on the dynamical behavior of the system.
\end{abstract}

\maketitle

\section{Introduction}

The investigations concerning dynamical properties of quantum impurity systems
are of great importance for the development of nanoscale
and, in general, condensed matter physics.
Precise control and manipulation of spin and charge degrees of freedom in such systems,
as well as understanding of relevant times scales,
is a necessary requirement for further applications in spintronics \cite{Wolf2001,DasSarma2004}
or for quantum information processing \cite{Deutsch1995,LossDiVincenzo1998}.
In addition, the analysis of dynamical behavior of various quantum impurity models
provides an important knowledge about the charge and spin
transport through nanostructures, and sheds new light
on the effects of decoherence and dissipation \cite{Menskii2003}.

A quantum impurity system can be regarded
as composed of a confined, zero-dimensional subsystem
interacting with infinitely large environment.
Recently, a prominent example undergoing vast theoretical and experimental
explorations is the system built of quantum dots or molecules attached to external leads.
Present nanofabrication techniques allow in particular for engineering
devices consisting of multiple quantum dots in various geometrical arrangements
and with precisely tuned parameters. This provides an unprecedented
opportunity for experimental investigations of many important effects present in such systems,
including the Kondo effect \cite{Goldhaber1998,Kouwenhoven1998},
superconducting correlations and Andreev transport \cite{Hofstetter2009, DeFranceschi2010},
quantum interference effects as well as various charge and spin transport
phenomena among many others \cite{Shtrikman1995, Donarini2019, Kouwenhoven1997},
and confront the experimental observations with the theoretical studies.

In addition to the examinations of the steady-state transport
properties of quantum dot systems,
there is an increasing number of experiments
conducted in the strong coupling regime,
where the dynamics and relaxation \cite{Loth2010, Terada2010, Yoshida2014, Cetina2016}
as well as different quench protocols and the Kondo physics have been investigated
in time domain \cite{Latta2011, Haupt2013}.
From theoretical point of view, the dynamical properties of low-dimensional systems
have been attracting a nondecreasing attention
\cite{Langreth1991,Rosch2003,White2004,Schiro2010,Gull2011,KennesRenorm2012}.
However, an accurate description of dynamics in such systems
poses a considerable challenge due to electronic correlations.
Recently, there have been significant advances in this regard
\cite{Schmidt2008,Vasseur2013,Antipov2016,Greplova2017,
FrohlingAnders2017,Maslova2017,Haughian2018,Moca2018,Domanski2018},
especially by resorting to various renormalization group schemes
\cite{CazalillaPRL2002,KirinoJPS2008,Tureci2011,Andergassen2011,
AO_andreas2012,Eidelstein2012,Kennes2012,Guttge2013,Kennes2014,
Weymann2015,Bidzhiev2017}.

In this paper, we address the problem of dynamical behavior
of quantum dots attached to spin-polarized leads
and focus on the strong coupling regime,
when electron correlations can give rise to the Kondo effect \cite{Kondo1964, Hewson1997}.
Perturbative approaches fail to capture strong correlations due to infrared divergences,
therefore, we turn to the Wilson's numerical renormalization group (NRG) method
\cite{Wilson1975} - a very accurate, non-perturbative method for calculating
transport properties of quantum impurity systems, including quantum dots coupled to
external leads. As we are interested in the charge and spin dynamics,
we use the extension of NRG introduced by Anders and Schiller,
namely the time-dependent numerical renormalization group (tNRG) method \cite{Anders2005, Anders2006}.
This method was subsequently generalized by Nghiem and Costi to finite temperatures,
multiple quenches and possibility to study time evolution in response to general pulses and periodic driving
\cite{Costi2014generalization, Costi2014, Costi2018}.
While tNRG has already provided a valuable insight into the dynamics
of Kondo-correlated molecules and quantum dots
attached to nonmagnetic leads \cite{Roosen2008, LechtenbergAnders2014, Weymann2015, Costi2017},
the time-dependent transport properties
of correlated impurities with spin-polarized contacts remain rather unexplored.
The goal of this paper is to fill this gap.

Quantum dots coupled to ferromagnetic electrodes
have already been extensively studied in the case
of stationary-state transport properties \cite{Martinek2003, Choi2004, Martinek2005, Sindel2007}.
In particular, the competition between the Kondo correlations and ferromagnetism
was shown to result in many nontrivial spin-related phenomena,
such as the exchange-field-induced suppression of the Kondo correlations
\cite{pasupathy_04,Matsubayashi2007,Hauptmann2008Mar,Gaass2011,weymannPRB11,Weymann2018KondoFM}.
Motivated by the above advances, we analyze the time-dependent properties
of a single quantum dot strongly coupled to ferromagnetic leads subject to a quantum quench.
More specifically, we consider two types of quantum quenches:
the first one concerns the quench in the spin-dependent coupling strength,
whereas the second type of quench is associated with a change in the dot's orbital level position.
We study the time evolution of the dot's magnetization and the occupation number following the quench.
Finally, we also take under consideration finite temperature effects
and analyze their impact on the spin dynamics.

We show that the time evolution of the dot's magnetization and occupation
strongly depends on the initial conditions of the system.
In particular, for the quantum dot initially occupied by a single electron,
we find a range of time where the time evolution of magnetization exhibits
a nonmonotonic behavior---magnetization shows
oscillations as a function of time with a sign change.
The corresponding sign change is also clearly
visible in the time dependence of the induced exchange field.
We show that this nonmonotonic buildup is a consequence
of qualitatively different time evolution of
spin-resolved occupations of the quantum dot.
It turns out that while the charge dynamics is mainly governed
by the coupling to majority spin subband of the ferromagnet,
the spin dynamics is determined by the coupling to the minority spin band.
Finally, we demonstrate that all these effects can be smeared out
by thermal fluctuations, once the inverse of temperature
becomes comparable with the time scale
when the interesting physics occurs.

This paper is structured as follows.
Section \ref{theoretical framework} consists of
the Hamiltonian description of the considered system,
the overview of the quench protocol and a summary of
the numerical renormalization group method used for
numerical calculations of time-dependent expectation values.
In Sec. \ref{results} we present the numerical results
and relevant analysis for the quenches in the coupling strength and orbital level position.
We also present and discuss the effects of finite temperature on dynamical behavior.
Finally, the work is concluded in Sec. \ref{conclusions}.

\section{Theoretical framework} \label{theoretical framework}
\subsection{Hamiltonian}

\begin{figure}[t]
  \includegraphics[width=.55\columnwidth]{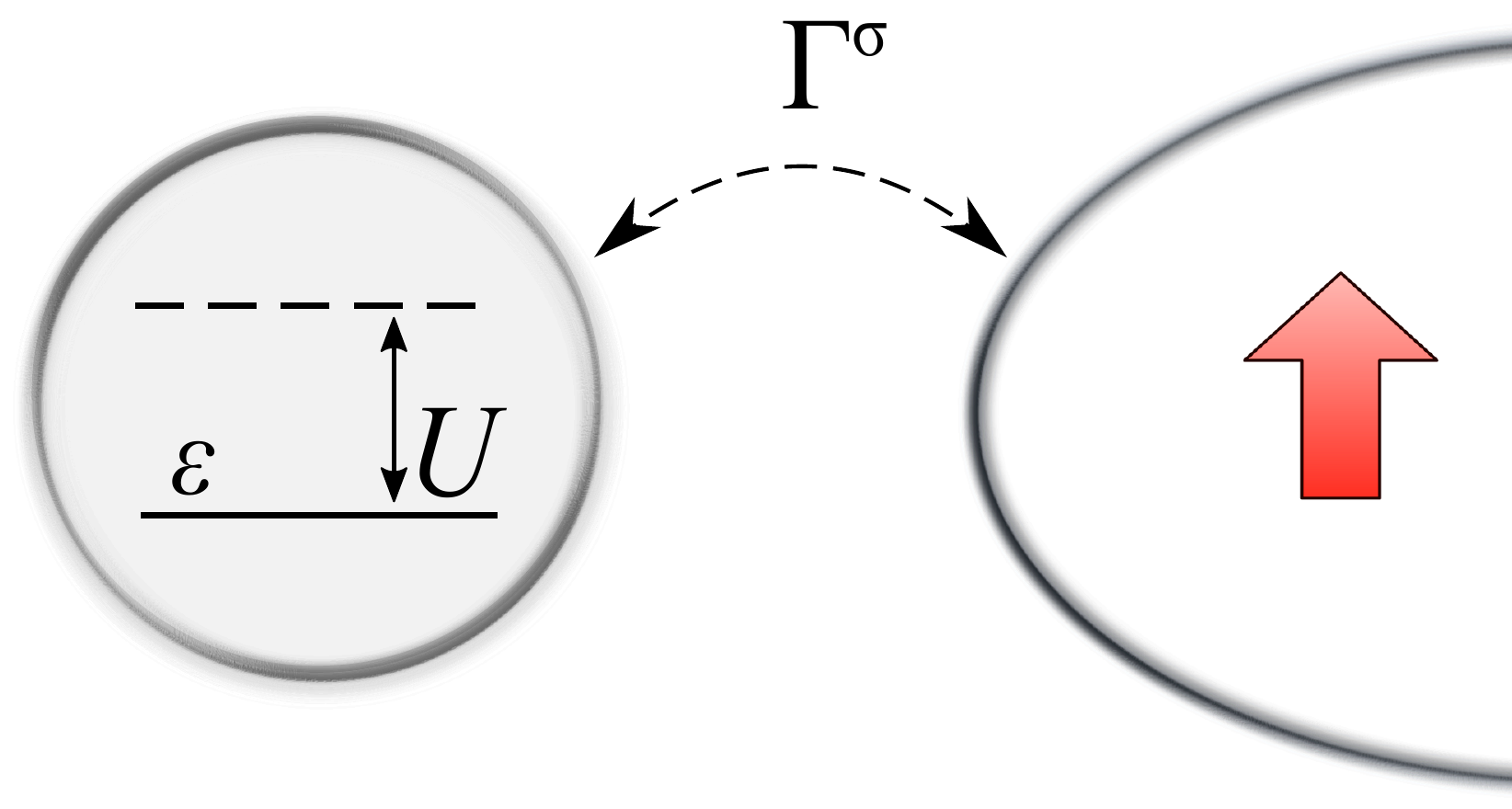}
  \caption{\label{Fig:1}
  Schematic of the considered system.
  A single-level quantum dot, with on-site energy $\varepsilon$
  and Coulomb correlations $U$, is attached to
  an effective reservoir of spin-polarized electrons
  with the spin-dependent coupling strength $\Gamma^\sigma$.
}
\end{figure}

We consider a single-level quantum dot
coupled to a spin-polarized ferromagnetic lead \cite{Martinek2003, Choi2004, Martinek2005, Sindel2007},
as shown in \fig{Fig:1}.
The system is described by the single-impurity Anderson Hamiltonian,
which can be generally written as
\begin{equation}\label{Eq:HamiltonianTotal}
H=H_{\mathrm{QD}}+H_{\mathrm{Lead}}+H_{\mathrm{Tun}}.
\end{equation}
The quantum dot Hamiltonian is given by
\begin{equation}\label{Eq:HamiltonianQD}
H_{\mathrm{QD}}=\varepsilon n + U n_{\uparrow}n_{\downarrow},
\end{equation}
where the quantum dot occupation is expressed as,
$n=n_\uparrow+n_\downarrow=d^\dagger_\uparrow d_\uparrow + d^\dagger_\downarrow d_\downarrow$,
with $d^\dagger_\sigma$($d_\sigma$) being the dot's fermionic creation (annihilation) operator
for an electron with spin $\sigma$ and energy $\e$.
The Coulomb correlation energy between the two electrons
occupying the dot is denoted by $U$.
The ferromagnetic lead is modeled as a reservoir of noninteracting quasiparticles,
\begin{equation}\label{Eq:HamiltonianLeads}
  H_{\mathrm{Lead}}=\sum_{\textbf{k}\sigma}\varepsilon_{\textbf{k}\sigma} c^\dagger_{\textbf{k}\sigma} c_{\textbf{k}\sigma},
\end{equation}
where $c^\dagger_{\textbf{k}\sigma}$($c_{\textbf{k}\sigma}$)
is the creation (annihilation) operator of an electron
with momentum $\textbf{k}$, spin $\sigma$ and energy $\varepsilon_{\textbf{k}\sigma}$.
Finally, the tunneling Hamiltonian reads
\begin{equation}\label{Eq:HamiltonianTun}
  H_{\mathrm{Tun}}=\sum_{\textbf{k} \sigma} V_{\sigma} (c^\dagger_{\textbf{k}\sigma}d_{\sigma}  + {\rm H.c.}),
\end{equation}
where the tunnel matrix elements are denoted by $V_{\sigma}$
and assumed to be momentum independent.
The spin-dependent coupling between the quantum dot and the lead
is expressed as, $\Gamma^\sigma = \pi \rho^\sigma |V_\sigma|^2$, with
$\rho^\sigma$ being the spin-dependent density of states of ferromagnetic electrode.
By introducing the spin polarization of the lead $p$,
the coupling strength can be written in the following manner,
$\Gamma^{\uparrow(\downarrow)}=\Gamma(1\pm p)$,
with $\Gamma^{\uparrow(\down)}$ denoting the coupling
to the spin-up (spin-down) electron band of the ferromagnetic lead
and $\Gamma= (\Gamma^\uparrow + \Gamma^\downarrow)/2$.

It is also worth of note that the considered model
is equivalent to a quantum dot coupled to the left and right leads
at equilibrium with the magnetic moments of the leads
forming a parallel alignment.
By performing an orthogonal transformation,
one can show that the quantum dot couples only to an even linear combination
of electrode's operators, with an effective coupling strength $\Gamma$
and average spin polarization $p$ \cite{Glazman1988}.

\subsection{Quench protocol}

In this paper the primary focus is put on understanding the spin-resolved
dynamics of the system subject to a quantum quench.
In general, the time-dependent Hamiltonian describing
the evolution after the quantum quench can be written as
\begin{equation}\label{Eq:Hamiltonian_TD}
  H(t) = \theta(-t)H_0 + \theta(t)H,
\end{equation}
where $\theta(t)$ is the step function.
Here, the Hamiltonian $H_0$ is the initial Hamiltonian of the system.
At time $t=0$, the system becomes quenched, i.e. its Hamiltonian
suddenly changes, and it evolves according to $H$.
The two Hamiltonians are thus given by \eq{Eq:HamiltonianTotal}
with appropriately changed parameters.
The time evolution of an expectation value of a local operator $\Op(t)$
can be then found from
\begin{eqnarray}\label{Eq:O}
  O(t) \equiv \langle \Op(t) \rangle = \mathrm{Tr}\left\{e^{-iHt} \rho_0 e^{iHt} \Op\right\},
\end{eqnarray}
where $\rho_0$ denotes the initial equilibrium density matrix
of the system described by $H_0$.

In the following, we study two types of quantum quenches.
In the first case, the quench concerns the coupling strength $\Gamma$.
It is assumed that for $t<0$, the quantum dot is decoupled from the lead ($\Gamma_0=0$)
and the quench takes place at $t=0$, suddenly changing Hamiltonian
from $H_0$ to $H$, with the spin-dependent coupling to ferromagnetic contact $\Gamma^\sigma$ being abruptly switched on.
The second type of quench that we investigate involves a change in the dot's orbital level position $\varepsilon_{0} \rightarrow \varepsilon$,
while the coupling strength remains intact.

For those two quenches we determine the time-dependence of expectation
values of the dot's magnetization and occupation.
The former one can be found from
\begin{equation}\label{Eq:Sz}
  S_z(t) = \frac{n_\uparrow(t)-n_\downarrow(t)}{2},
\end{equation}
which can be easily expressed with the use of quantum dot's operators as,
$n_\uparrow(t)=d^\dagger_\uparrow(t) d_\uparrow(t)$ and
$n_\downarrow(t)=d^\dagger_\downarrow(t) d_\downarrow(t)$,
whereas the latter one is simply equal to $n(t) = n_\up(t) + n_\down(t)$.

\subsection{NRG implementation}

To account for various many-body effects and analyze
the spin-resolved dynamics in most accurate manner,
we use the Wilson's numerical renormalization group method \cite{Wilson1975, Bulla2008}
to find the eigenspectrum of the Hamiltonian (\ref{Eq:HamiltonianTotal}).
At first, the conduction band of the lead is logarithmically
discretized with a discretization parameter $\Lambda$.
Consequently, the discretized band is mapped on a tight-binding chain with
exponentially decaying hopping between the consecutive sites, forming the Wilson chain \cite{Bulla2008}.
After this transformation the Hamiltonian (\ref{Eq:HamiltonianTotal}) can be explicitly written as
\begin{equation}\label{Eq:HamiltonianNRG}
\begin{split}
H=&H_{\mathrm{QD}} + \sum_\sigma V_\sigma(f^\dagger_{0\sigma} d_\sigma + d^\dagger_\sigma f_{0 \sigma})\\
 +&\sum_{n=0}^\infty \sum_{\sigma}\xi_n(f^\dagger_{n\sigma}f_{n+1\sigma} +f^\dagger_{n+1\sigma} f_{n\sigma}) .
\end{split}
\end{equation}
Here, the operator $f^\dagger_{n \sigma}$ creates an electron of spin-$\sigma$
at the $n$th site of the Wilson chain,
while $\xi_n$ denotes the hopping integrals between
the sites $n$ and $n+1$, respectively \cite{Wilson1975, Bulla2008}.
The Hamiltonian $H_0$ is also given by \eq{Eq:HamiltonianNRG}
with appropriately adjusted parameters.

We diagonalize both Hamiltonians, $H$ and $H_0$, using NRG \cite{NRG_code}
in $N$ iterations and keeping up to $N_K$ energetically lowest-lying eigenstates
retained at each iteration of the NRG procedure.
These states are referred to as {\it kept} and
labeled with the superscript $K$.
For a few first sites of the Wilson chain, $n<n_0$, all the states are kept.
However, once the size of the Hilbert space exceeds $N_K$, which happens
at certain iteration $n=n_0$, one needs to truncate the space
by discarding the high-energy eigenstates.
These states are referred to as {\it discarded}
and labeled with the superscript $D$.
In addition, all the states of the last iteration $n=N$
are also considered as discarded states.

The discarded states $|ns\rangle^{D}$
at iterations $n<N$ are complemented by the state space of the rest
of the chain spanned by the environmental states $|ne\rangle$ \cite{Anders2005, Anders2006}.
The resulting states
\begin{equation}
|nse\rangle^{D} \equiv  |ns\rangle^{D} \otimes |ne\rangle,
\end{equation}
allow us to find the full many-body eigenbases
\begin{equation} \label{eq:completeness}
\sum_{nse}|nse\rangle^{\!D}_{0} \,{}^{D}_{\,0}\!\langle nse| \!=\! \mathbbm{1} \;\;\;\,
 {\rm and}
 \,\;\;\; \sum_{nse}|nse\rangle^{\!D} \,{}^D \!\langle nse| \!=\! \mathbbm{1},
\end{equation}
of the two Hamiltonians, $H_0$ and $H$, respectively.
Here, the summation over the Wilson shells $n$
involves only the shells where discarded states are designated,
i.e. ${ \sum_n \equiv \sum^N_{n \geqslant  n_0}}$.
The above eigenbases, due to the energy-scale separation,
are good approximations of the eigenstates of the full NRG Hamiltonians
\cite{Anders2005, Anders2006}
\begin{eqnarray}
H_{0}|nse\rangle^X_0 &\simeq& E_{0ns}^X |nse\rangle^X_0,\\
H|nse\rangle^X &\simeq& E_{ns}^X|nse\rangle^X,
\end{eqnarray}
where $X=K$ ($X=D$) denotes a kept or a discarded state.

The discarded states of the Hamiltonian $H_0$ are
furthermore used to construct the full density matrix of the system
at temperature $T\equiv1/\beta$ \cite{Andreas_broadening2007}
\begin{equation}
\rho_0=\sum_{nse}\frac{e^{-\beta E_{0ns}^D}}{Z} |nse\rangle^D_0{}^D_0\langle nse|,
\end{equation}
where
\begin{equation}
Z\equiv\sum_{nse} e^{-\beta E_{0ns}^D}
\end{equation}
is the partition function.
Note that the energies $E_{0ns}^D$ are independent of the environmental index $e$.
Tracing out the environmental states introduces
the weight factor $w_n\equiv\frac{d^{N-n}Z_n}{Z}$ of a given iteration
\cite{Andreas_broadening2007}
\begin{equation}
\rho_0=\sum_n \underbrace{\frac{d^{N-n}Z_n}{Z}}_{w_n} \underbrace{\sum_{s}\frac{e^{-\beta E_{0ns}^D}}{Z_n} |ns\rangle^D_0{}^D_0\langle ns|}_{\text{$\rho_{0n}$}},
\end{equation}
with
\begin{equation}
Z_n\equiv \sum_{s} e^{-\beta E_{0ns}^D}
\end{equation}
denoting the partition function of a given iteration
and $d$ being the local dimension of the Wilson site.
Consequently, the density matrix can be written in a compact form as
\begin{equation}\label{eq:rho0n}
\rho_0=\sum_{n}w_n \rho_{0n}.
\end{equation}

The time-dependent expectation value $\langle \Op(t) \rangle$ of an operator $\Op$
[cf. \eq{Eq:O}] can be written using the complete NRG bases as
\begin{eqnarray}\label{Eq:Ot1}
    \langle \Op(t) \rangle &=&\!\!
    \sum_{nn'n''} \sum_{ss'e}  {}^{ D}\! \langle nse|w_{n''} \rho_{0n''}| n's'e\rangle^{\! D} \nonumber \\
   &&\times {}^{ D}\! \langle n's'e|\Op|nse\rangle^{\! D} \; e^{i(E^{ D}_{ns} - E^{ D}_{n's'})t}.
\end{eqnarray}
Note that this formula involves a triple summation over the Wilson shells,
one summation results from the definition of the full density matrix
$\rho_0$ [cf. \eq{eq:rho0n}], whereas the two other stem
from the completeness relation [cf. \eq{eq:completeness}].
To make this formula computationally more efficient,
such that one could make the calculations in a
single-sweep fashion, we use the identity
$\sum_{nn'} \equiv \sum_{n}^{ XX'\neq KK}$,
in which the double sum over the states of the Wilson chain
is changed into a single sum over $n$
with an additional summation over the combination
of kept and discarded states, except when both states are kept
\cite{Weymann2015}.
Then, the formula for the expectation value, \eq{Eq:Ot1}, becomes
\begin{eqnarray}\label{Eq:Ot2}
    \langle \Op(t) \rangle &=&\!\!
    \sum_{n}^{XX'\neq KK}\sum_{n'} \sum_{ss'e}  {}^{ X}\! \langle nse|w_{n'} \rho_{0n'}| ns'e\rangle^{\! X'} \nonumber \\
   &&\times {}^{ X'}\! \langle ns'e|\Op|nse\rangle^{\! X} \; e^{i(E^{ X}_{ns} - E^{ X'}_{ns'})t}.
\end{eqnarray}
This formula can be directly evaluated by using NRG in time-domain \cite{Costi2014},
however, it is more convenient to perform the time-dependent calculations
in the frequency space and then apply the Fourier transformation
back to the time domain \cite{Andreas2012}.
The frequency-dependent expectation value $\langle \Op (\omega) \rangle$
of a local operator $\Op$ is given by
\begin{eqnarray}\label{Eq:Ow}
    \langle \Op(\omega) \rangle &=&\!\!
   \sum_{n}^{ XX'\neq KK}\sum_{n'} \sum_{ss'e}  {}^{X}\! \langle nse|w_{n'} \rho_{0n'}| ns'e\rangle^{\! X'} \nonumber\\
   &&\times {}^{ X'}\! \langle ns'e|\Op|nse\rangle^{\! X} \; \delta(\omega + E_{ns}^{X} - E_{ns'}^{X'}).
\end{eqnarray}
It is interesting to note that the calculations of the frequency-dependent expectation value
can be performed in a similar fashion to the calculation of the spectral function
within conventional NRG \cite{Costi1997,Bulla2008, Andreas_broadening2007, Toth2008}.

\begin{figure}[t]
  \includegraphics[width=.7\columnwidth]{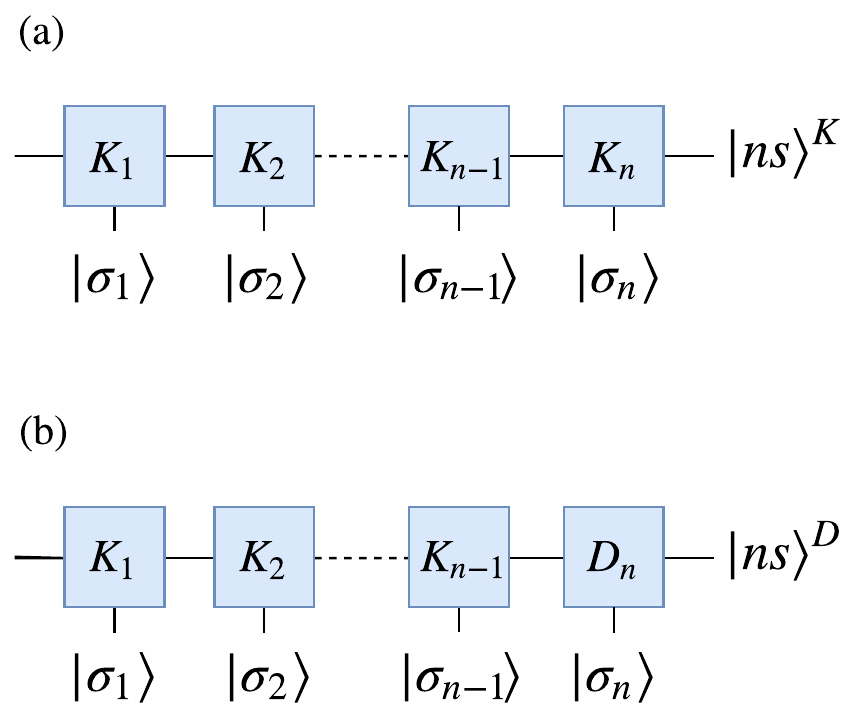}
  \caption{\label{Fig:MPS}
  Matrix product illustration of (a) kept ($\ket{ns}^K$)
  and (b) discarded ($\ket{ns}^D$) state at Wilson shell $n$.
  The bottom legs label the local states $|\sigma_n\rangle$.
  The blocks $K_n$ ($D_n$) represent the kept (discarded)
  state space at Wilson shell $n$.
}
\end{figure}

\subsection{Calculation procedure}

\begin{figure}[t!]
  \includegraphics[width=.95\columnwidth]{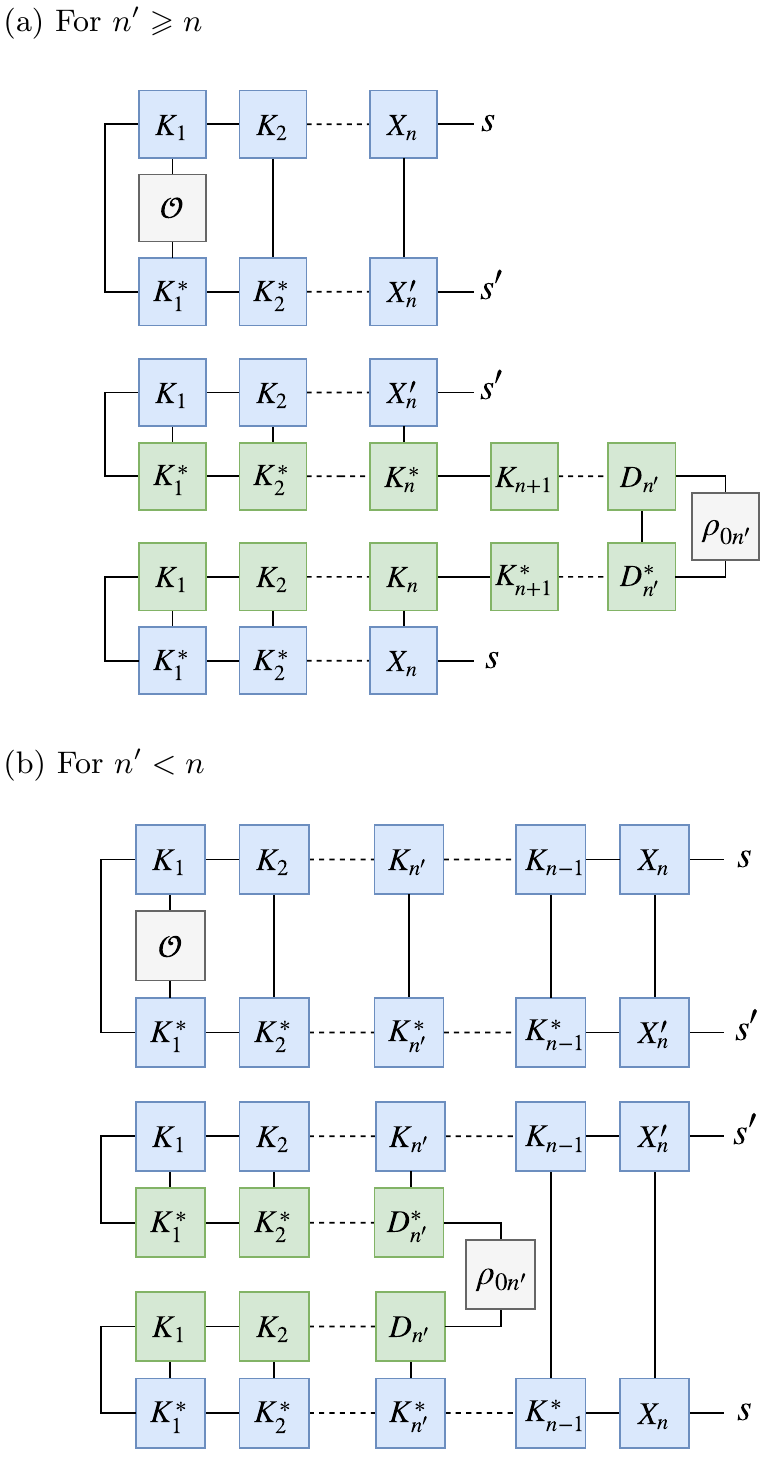}
  \vspace{-0.3cm}
  \caption{\label{Fig:2}
  Matrix product state diagrams for the calculation of a contribution
  to the frequency-dependent expectation value of
  an operator $\Op$ after the quantum quench, as given by Eq. (\ref{Eq:Ow}).
  The first diagram (a) shows the contribution
  relevant for ${n' \geq n}$, whereas
  the second diagram displayed in (b) presents the contribution
  for ${n' < n}$.
  These contributions need to be summed over the states
  $s$ and $s'$ and iterations $n$ and $n'$.
  For the contribution presented in (b) there
  is an additional weight factor given by $d^{n'-n}$
  due to the environmental states.
  The green squares represent the state space of the initial Hamiltonian $H_0$,
  whereas the blue squares represent the states of
  the final Hamiltonian $H$.
}
  \vspace{-0.3cm}
\end{figure}

All the calculations can be conveniently performed
in the matrix product states language \cite{Verstraete2009, Schollwock2011, Andreas2012}.
An exemplary illustration of a kept or a discarded state $\ket{ns}^X$ is presented
in \fig{Fig:MPS}. Using MPS diagrammatics,
the frequency-dependent expectation value of an operator $\Op$
given by \eq{Eq:Ow}
can be calculated in an iterative fashion,
where the data points corresponding to $\omega = E_{ns'}^{X'} - E_{ns}^{X}$
are collected in appropriate energy bins on logarithmic scale.
The part of the expression for $\langle \Op(\omega) \rangle $
preceding the Dirac delta function can be estimated from the
MPS diagrams shown in Fig. \ref{Fig:2}.
In calculations, it is important to consider separately
the case of $n' \geq n$ and $n'< n$,
depending on whether the density matrix $\rho_0$ gives
the contribution at iterations equal or larger than $n$ or smaller than $n$.
In the first situation, one needs to evaluate the
MPS diagrams shown in \fig{Fig:2}(a).
On the other hand, in the second case of $n'< n$,
the corresponding diagram is illustrated in \fig{Fig:2}(b).
Note that in this situation, the trace over the environmental
states results in a weight factor given by $d^{n'-n}$.
Notice also that at $T=0$, i.e. for the ground state,
only the first MPS diagram, which is shown in \fig{Fig:2}(a), is relevant.
All these contributions need to be summed over
the states and the Wilson shells, as given explicitly in \eq{Eq:Ow}.
Eventually, one obtains the spectral representation
of an expectation value of $\Op(t)$
given by a sum of Dirac delta peaks with the corresponding weights $\Op_j$
\begin{equation}
   O(\omega)\equiv \langle \Op(\w) \rangle = \sum_j \Op_j \delta(\w-\w_j).
\end{equation}
The delta peaks consist of one large contribution
at $\omega \to 0$, which corresponds to the long-time-limit value of $O(t)$.
The collected delta peaks are then log-Gaussian broadened
with a broadening parameter $b$ (except for
the point at $\omega \to 0$)
and Fourier-transformed back into the time domain to finally obtain
\begin{equation}\label{Eq:Ot}
  O(t) = \int_{-\infty}^\infty \!{O(\omega) e^{-i\omega t} d \omega}.
\end{equation}

As far as NRG technicalities are concerned,
in calculations we assumed the discretization parameter $\Lambda=2$,
set the length of the Wilson chain to be $N=80$
and kept at least $N_K=2000$ energetically lowest-lying eigenstates at each iteration.
Moreover, to increase the accuracy of the data
and suppress the band discretization effects,
we employ the Oliveira's $z$-averaging \cite{Oliveira1990}
by performing calculations for $N_z=8$ different discretizations.

\begin{figure}[t]
  \includegraphics[width=.9\columnwidth]{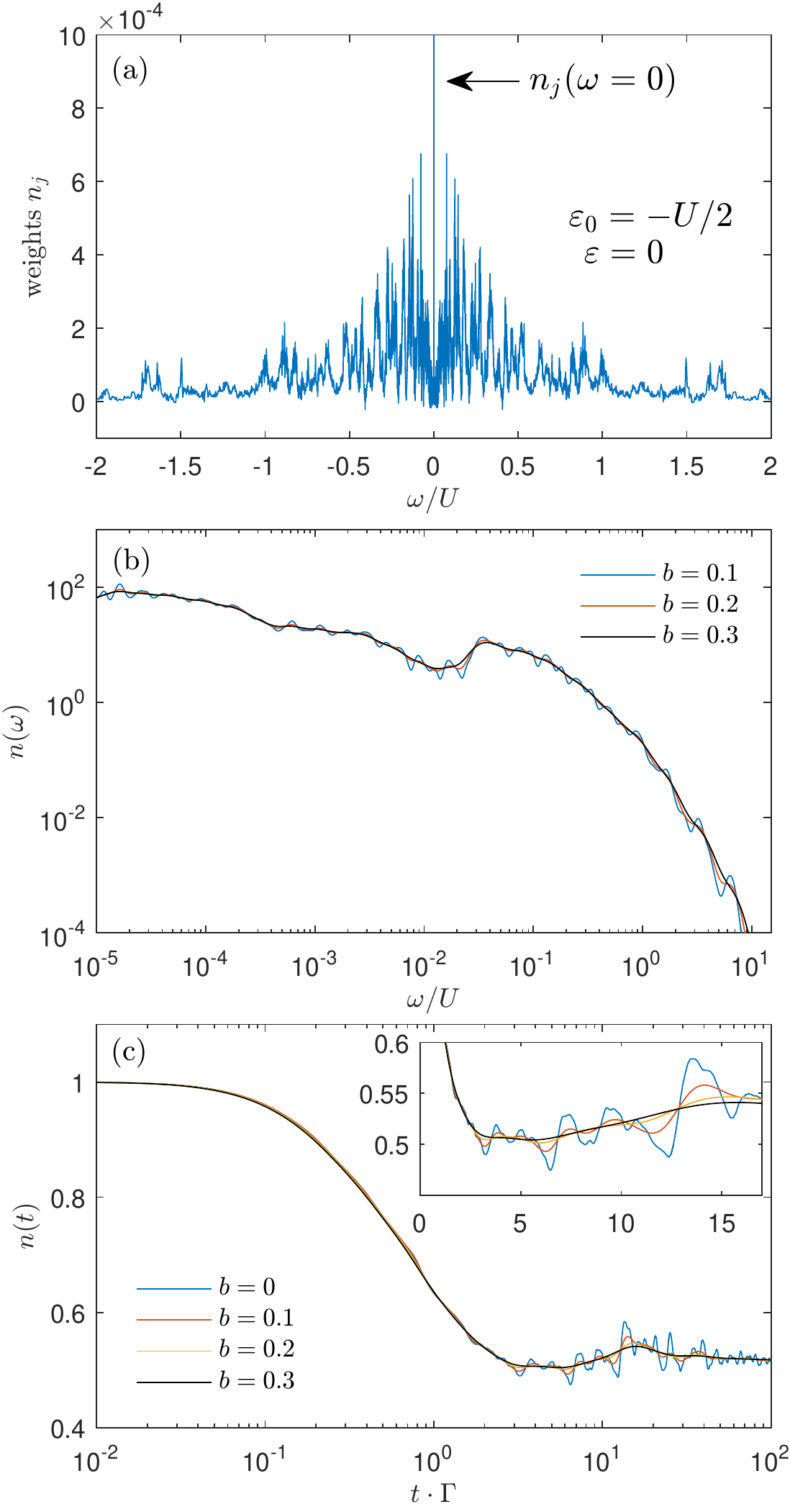}
  \vspace{-0.3cm}
  \caption{\label{Fig:4}
  The results for quench preformed in the dot's level position
  from $\varepsilon_0=-U/2$ to $\varepsilon=0$.
  Panel (a) presents the weights $n_j$ of collected delta peaks
  corresponding to the frequency dependent local operator
  $n(\omega)=\sum_j n_j \delta(\omega - \omega_j)$.
  Panel (b) shows the absolute value of the collected
  delta peaks after the logarithmic-Gaussian broadening
  (without the point at $\omega \to 0$) for different values of the
  broadening parameter $b$, as indicated,
  plotted as a function of frequency $\omega$ on the logarithmic scale.
  The time-dependence of $n(t)$ for the considered quench is shown in (c).
  The inset in (c) presents the dependence of $n(t)$ in the linear scale.
  The parameters are: $T=0$, $U=0.12$ (in units of band halfwidth),
  $p=0.5$, and $\Lambda=2$, $N_K=2000$, $N_z=8$.
  \vspace{-0.3cm}
  }
\end{figure}

\begin{figure}[t]
  \includegraphics[width=.9\columnwidth]{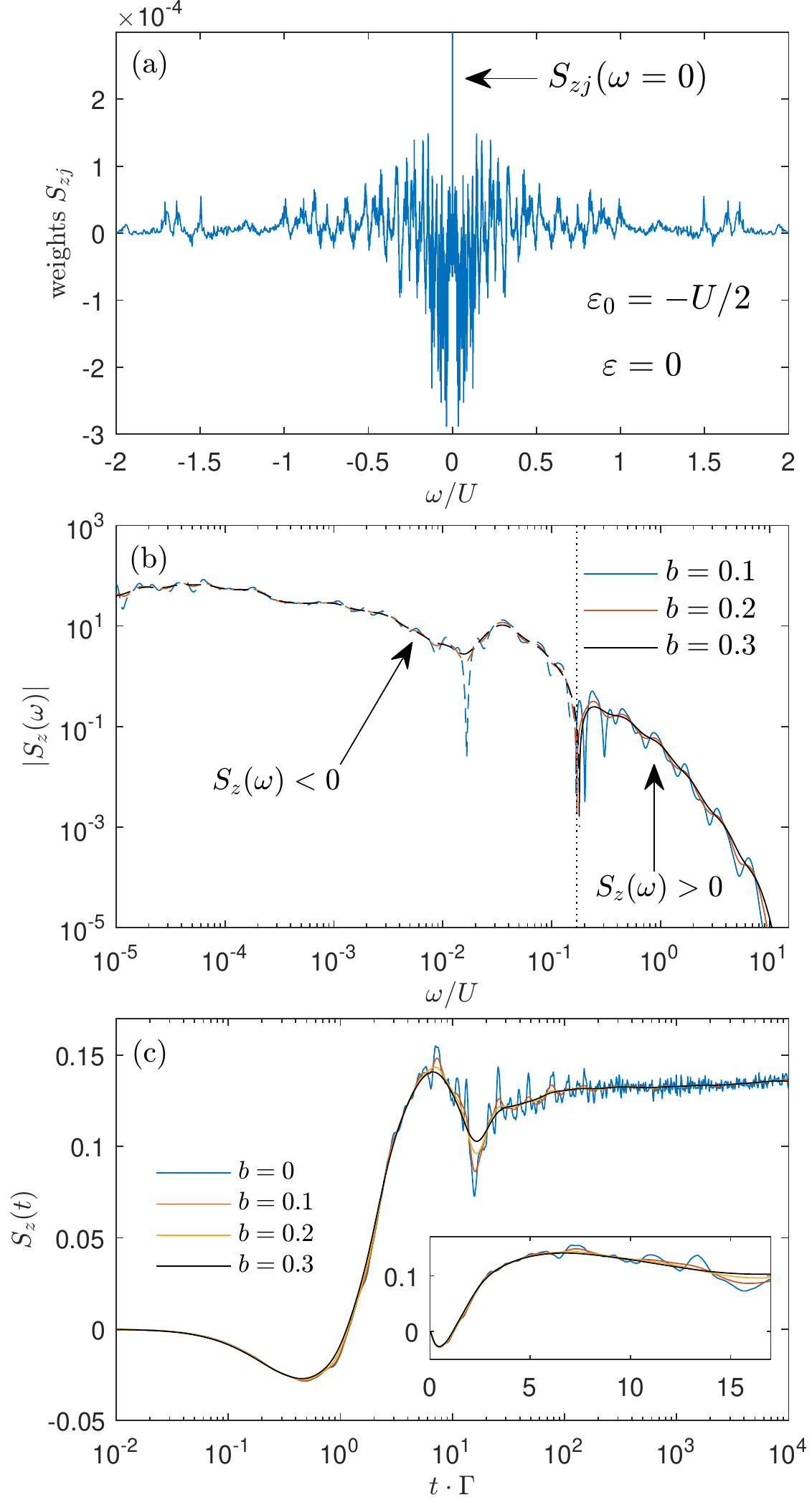}
  \caption{\label{Fig:4a}
  The same as in \fig{Fig:4} calculated for the dot's magnetization $S_z(t)$.
  The vertical dotted line in (b) marks the frequency
  at which the spectral density $S_z(\omega)$ changes sign:
  For $\omega/U\gtrsim 0.2$, $S_z(\omega)>0$ (solid lines),
  while for $\omega/U\lesssim 0.2$, $S_z(\omega)<0$ (dashed lines).}
\end{figure}

In Figs.~\ref{Fig:4} and \ref{Fig:4a} we show exemplary results
for the quantum dot occupation number and magnetization,
respectively, obtained for a quench performed in the dot's level position.
The initial Hamiltonian $H_0$ has the orbital level set to
the particle-hole symmetry point, $\varepsilon_0=-U/2$,
while for the final Hamiltonian $H$ the level is set at resonance $\varepsilon=0$.
The collected delta peaks obtained from the calculations
along with their weights, cf. Eq.~(\ref{Eq:Ow}), are shown
in the top panels of Figs.~\ref{Fig:4} and \ref{Fig:4a}.
The black arrows at $\omega=0$ indicate
the zero energy peak corresponding to the long-time-limit value of the
corresponding expectation value.
In the next step, the delta peaks are broadened using
the logarithmic Gaussian kernel with the broadening parameter $b$ \cite{Andreas_broadening2007}.
The broadened data is presented in Figs.~\ref{Fig:4}(b) and \ref{Fig:4a}(b)
for different values of the broadening parameter.
It can be seen that with increasing the value of $b$
the artifacts resulting from discretization of conduction band
become averaged out. The broadened data is subsequently
Fourier-transformed to obtain the time-dependent expectation value.
The time evolution of the dot's occupation number and magnetization
is presented in Figs.~\ref{Fig:4}(c) and \ref{Fig:4a}(c),
correspondingly, for a few selected values of $b$.
Note that the case of $b=0$ corresponds to
obtaining $O(t)$ directly from discrete data without broadening.
However, to suppress the discretization artefacts and obtain smooth
data, in the next sections we use the broadening parameter equal $b=0.3$.


\section{Results and discussion} \label{results}


In the following, we present and discuss the
behavior of the dot's magnetization and occupation
as a function of time considering quenches both in the coupling
strength and the position of the dot's orbital level.
This allows us to investigate the build-up of the
exchange field in the system and study its dependence
on the model parameters and temperature.

\subsection{Quench in the coupling strength}


In this section we consider an initially ($t<0$) unpolarized quantum dot decoupled from the lead,
i.e., $S_z(t<0)=0$ and $\Gamma_0=0$. The initial occupation number depends only on the
position of the dot's energy level $\varepsilon$,
which in experimental setup can be tuned by changing the electrostatic potential of the corresponding gate.
At time $t=0$, the coupling $\Gamma$ between the quantum dot and ferromagnetic lead
is abruptly switched on. Because of that, the spin-resolved charge fluctuations between the dot
and the lead become allowed, resulting in a spin-dependent renormalization
of the quantum dot level, which gives rise to its finite magnetization.

\subsubsection{Quantum dot's magnetization}

\begin{figure}[t]
  \includegraphics[width=1\columnwidth]{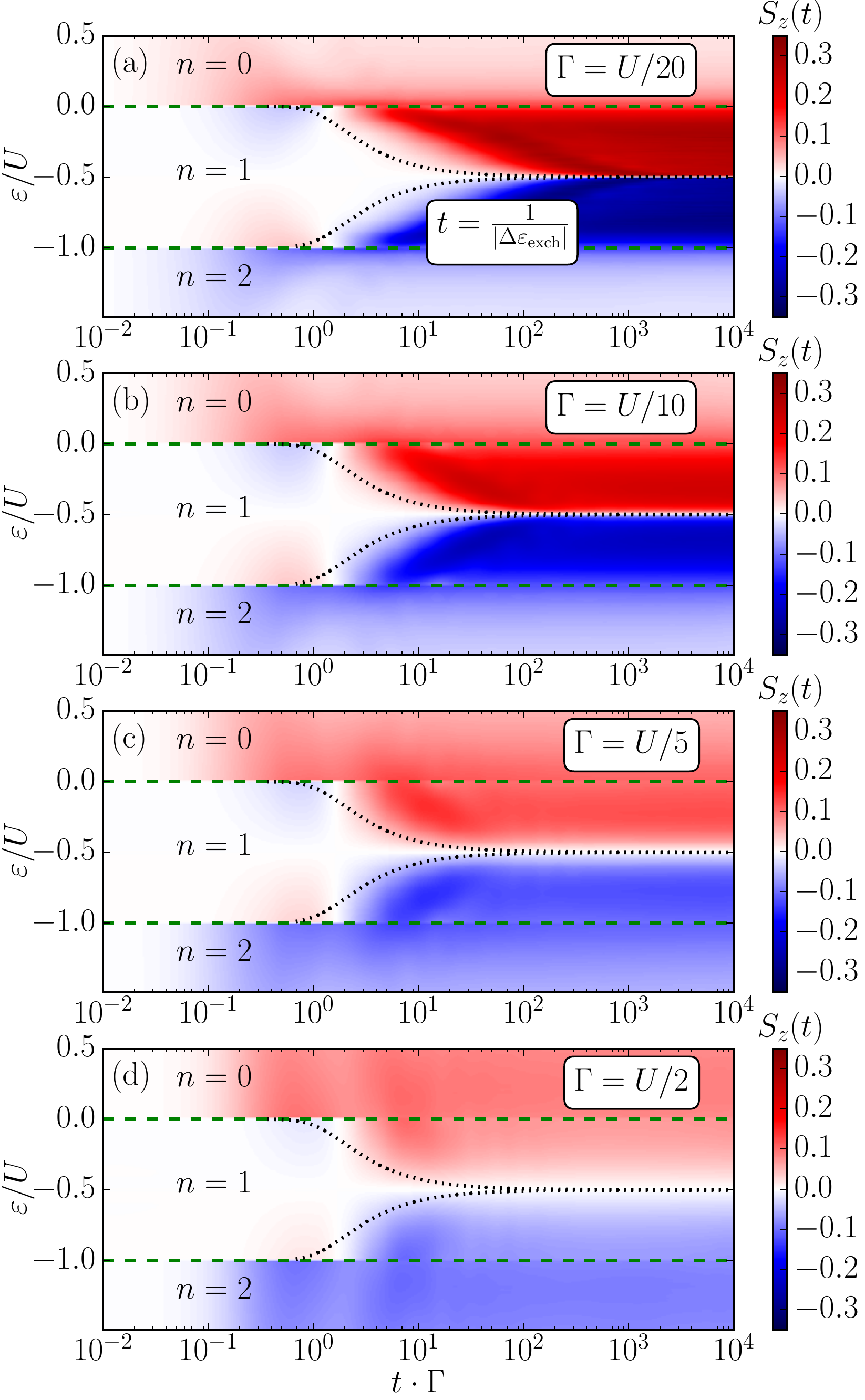}
  \caption{\label{Fig:5}
  The quantum dot magnetization $S_z(t)$, after the quench
  from an isolated and spin-unpolarized quantum dot
  to the coupled regime, as a function of time and the dots' energy level $\varepsilon/U$
  for different values of the coupling strength $\Gamma$, as indicated.
  The horizontal green-dashed lines separate regimes with different initial occupation number $n$.
  The black dotted lines indicate the time scale associated with the development of the exchange field $t = 1/|\exch|$.
  The calculations were performed for $T=0$, $U=0.12$ (in units of band halfwidth),
  $p=0.5$ and the following NRG parameters:
  $\Lambda=2$, $N_K=2000$, $N_z=8$ and $b=0.3$.
}
\end{figure}

The quantum dot magnetization $S_z(t)$
as a function of time and the position of the dot's energy level $\varepsilon$
is shown in Fig. \ref{Fig:5} for a few values of the coupling strength $\Gamma$.
The time evolution is calculated for a wide range
of position of the dot's energy level,
$-1.5 \leqslant \varepsilon/U \leqslant 0.5$,
therefore we are able to analyze in the full parameter space
how the initial occupation of the quantum dot influences
the spin dynamics after the quench in the coupling.
In general, one can clearly distinguish three regimes
with the quantum dot initially occupied by zero ($n=0$), one ($n=1$) and two (n=2) electrons.
The different occupation regimes are correspondingly
indicated and separated with dashed lines in Fig. \ref{Fig:5}.
Clearly, $S_z(t=0) = 0$ for all dot occupations,
since finite magnetization can build up
only due to spin-resolved fluctuations between the dot and ferromagnetic reservoir.
Thus, one should expect $S_z(t>0)\neq 0$.
With even number of electrons occupying
the quantum dot in the initial state,
the time-dependent magnetization $S_z(t)$ develops in time
acquiring only positive values for $n=0$ ($\e>0$) and negative values
for $n=2$ ($\e<-U$) occupation numbers.
Except for the opposite sign (direction of the magnetization),
the time evolution of magnetization is identical in both occupation regimes.
Apparently, when the quantum dot is either empty or doubly occupied,
the growth of the magnetization should not be possible.
However, finite coupling $\Gamma$ renormalizes and broadens
the dot's energy level, which for the initially empty quantum dot results in
a small growth of occupation $n(t>0)>0$, while for the initially doubly occupied dot
gives rise to the corresponding decrease of the double occupation $n(t>0)<2$.
Moreover, the spin-dependence of the coupling strength
lifts the degeneracy of singly occupied states,
which in consequence leads to a finite magnetization of the quantum dot.

This nonzero magnetization is a direct manifestation
of the so-called exchange field that builds up in the quantum dot
coupled to a reservoir of the spin-polarized electrons \cite{Martinek2003}.
The exchange field can be defined as
$\exch = \delta\e_\uparrow - \delta\e_\downarrow$,
where $\delta\e_\s$ is the renormalization of the spin-$\s$
dot level caused by the spin-dependent charge fluctuations.
The renormalization can be estimated within the second-order perturbation theory as
\cite{Martinek2003,Choi2004,Martinek2005,Sindel2007}
\begin{equation} \label{Eq:exch}
\exch = \frac{2 p \Gamma}{\pi} {\rm Re} \left[\phi(\e) - \phi(\e+U) \right],
\end{equation}
with $\phi(\e) = \Psi(1/2+i\e/2\pi T)$, where $\Psi(z)$ is the digamma function.
At zero temperature, the formula for the exchange field simply becomes
${\exch = \frac{2 p \Gamma}{\pi} \mathrm{ln} \left| \frac{\varepsilon}{\varepsilon+U} \right|}$.
Now, it can be clearly seen that $\exch$ changes sign exactly
at the particle-hole symmetry point, $\e=-U/2$. Consequently,
for $\e>-U/2$ ($\e<-U/2$) one finds $\exch<0$ ($\exch>0$),
which immediately implies that $S_z(t)>0$ [$S_z(t)<0$]
in the corresponding transport regime.
This behavior is clearly visible in \fig{Fig:5} in the even dot occupation regimes.

Let us now consider the most interesting transport regime
where initially the quantum dot is occupied by a single electron, i.e.
for $-U<\varepsilon<0$. As can be seen,
the general tendency of the behavior of $S_z(t)$ in the long-time limit
is consistent with the behavior of the exchange field discussed above.
Exactly at the particle-hole symmetry point
the charge fluctuations are the same for both spin directions
such that $\delta\e_\uparrow = \delta\e_\downarrow$
and $\exch = 0$ \cite{Martinek2003,Choi2004}.
This is why for $\e=-U/2$ the magnetization does not develop
and the dot remains unpolarized irrespective of time evolution, $S_z(t) = 0$.
However, when the energy of the orbital level is moved away
from the particle-hole symmetry point,
the time evolution of magnetization $S_z(t)$ shows a qualitatively different dependence.
For shorter times, ${0.1\lesssim t\cdot \Gamma \lesssim 1}$,
the magnetization points in the direction opposite to its long-time-limit value.
Around $t \approx 1/\Gamma$,
the sign change of magnetization occurs and
subsequently $S_z(t)$ grows and saturates at longer times, see \fig{Fig:5}.
One could expect that the time scale for the development
of the dot's magnetization (the exchange field) is simply given by
$t \sim 1/\exch$. This is however not entirely correct.
We would like to point out that the estimation of the
magnetization development time scale simply by $t = 1/\exch$
(see the black dotted lines in all panels of Fig. \ref{Fig:5})
does not fit to the numerically calculated dependence.
It is clearly visible that the dynamics of the exchange field development
is strongly influenced by the coupling strength
and does not scale linearly with $\Gamma$.

The comparison of the results for $S_z(t)$ when the coupling
strength $\Gamma$ is varied brings further important observations.
In the empty or doubly occupied dot regime, the magnitude of magnetization
becomes enhanced with increasing the coupling strength.
This is associated with an increase of level broadening and renormalization
effects as $\Gamma$ is increased. These effects enlarge
the occupation of the odd-electron states, which is responsible
for enhancement of $|S_z(t)|$, see \fig{Fig:5}. However,
as the occupation of odd-electron states becomes enhanced
in the even valleys, the same happens for even-occupation
states in the odd-electron valley. More precisely,
in the singly occupied dot regime,
as the coupling strength increases, the occupation of even-electron states
becomes enhanced at the cost of the odd states. Consequently,
in this transport regime one observes an opposite effect,
i.e. the larger becomes the coupling, the smaller the magnetization that develops in time is.

In addition, in the strong coupling regime also the Kondo correlations come into play.
Their role is reflected in the fact that now one needs to
detune the dot level more from the particle-hole symmetry point
to obtain a considerable magnetization.
As known from the studies of equilibrium transport properties of quantum dots
\cite{Martinek2003,Choi2004,Martinek2005,Sindel2007,weymannPRB11},
the Kondo resonance becomes suppressed when detuning
from the particle-hole symmetry point becomes so large
that the following condition is fulfilled $|\exch|\gtrsim T_K$,
where $T_K$ is the Kondo temperature.
This fact has also strong consequences for the dynamical behavior
of the system. Finite values of $S_z(t)$ develop
only when the above inequality becomes satisfied,
as otherwise the spin of the dot forms a delocalized
singlet state with conduction electrons
and the magnetization does not develop.

It is important to note that the variation of the coupling strength has also an important impact
on the corresponding time scales for the development of the dot's magnetization.
For smaller values of the coupling, see \fig{Fig:5}(a),
it takes longer time for the magnetization to fully develop,
whereas for stronger couplings this time scale becomes reduced, see \fig{Fig:5}(d).

\subsubsection{Buildup of exchange field}

Let us now focus on the time scales associated with the development
of dot's magnetization and the associated exchange field.
To estimate the magnitude of the exchange field
we compare the value of the time-dependent magnetization
to the static magnetization of a similar system coupled
to normal metallic leads in the presence of an external magnetic field $B$,
i.e. $S_z(t)=\langle S_z(B)\rangle$.
In order to solve this model at equilibrium,
we assume vanishing spin polarization of the leads $p=0$
and add the Zeeman energy term $H_\mathrm{B}=g \mu_\mathrm{B} B S_z$
to the quantum dot Hamiltonian $H_{\mathrm{QD}}$, with $g\mu_{\mathrm{B}}\equiv 1$.
We associate the Zeeman energy that results in magnetization
$\langle S_z(B)\rangle = S_z(t)$ with the exchange field energy $\exchn$.
In this manner, we are able to evaluate the time dependence
of the generated exchange field $\exchn(t)$.

\begin{figure*}[t]
  \includegraphics[width=1.85\columnwidth]{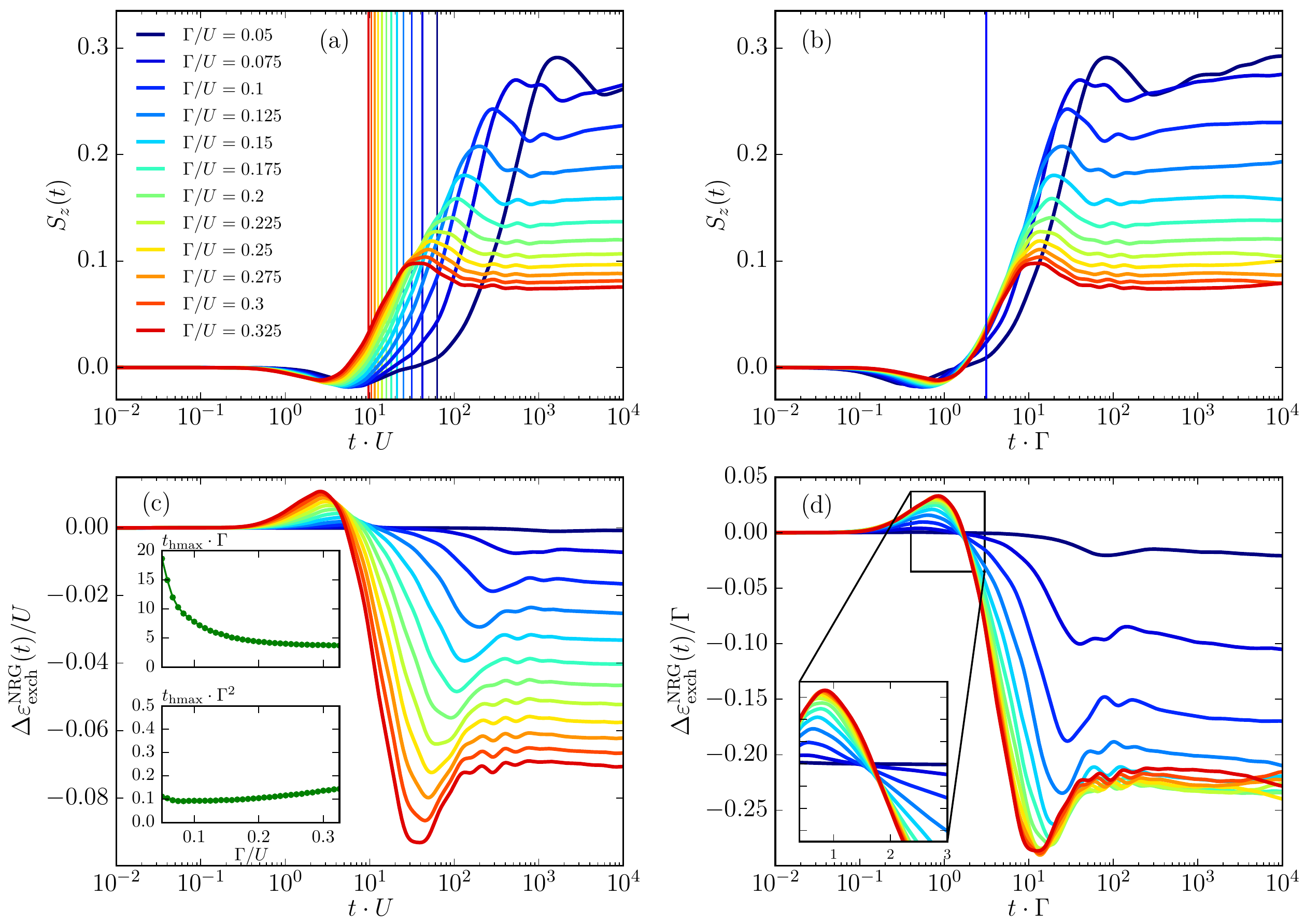}
  \caption{\label{Fig:6}
  (top row) The quantum dot magnetization $S_z(t)$
  and (bottom row) the induced exchange field $\exchn(t)$
  after the quench in the coupling strength plotted as a function of time
  calculated for different couplings, as indicated.
  The vertical lines in (a) and (b) indicate
  the time scale associated with the exchange field, $t = 1/\exch$, cf. \eq{Eq:exch}.
  The inset in (c) presents the time scale $t_{\rm hmax}$
  associated with the half-maximum value of $|\exchn(t)|$
  plotted as a function of $\Gamma$, whereas the inset in (d)
  shows a close-up of $\exchn(t)$
  for times where the sign change of exchange field occurs.
  The parameters are the same as in \fig{Fig:5}
  with quantum dot's energy level equal to $\varepsilon=-U/4$.
}
\end{figure*}

Figure \ref{Fig:6} presents the time evolution of the magnetization
[Figs. \ref{Fig:6}(a) and (b)] together with the evaluated
exchange field [Figs. \ref{Fig:6}(c) and (d)].
In this figure the energy of the orbital level is set to
$\varepsilon=-U/4$, corresponding to the transport regime
where finite magnetization develops and its sign change as the time elapses is visible,
cf. \fig{Fig:5}. In order to get information about the relevant time scales
we plot the dependence of the quantities of interest versus $t\cdot U$
[Figs. \ref{Fig:6}(a) and (c)] and $t\cdot\Gamma$ [Figs. \ref{Fig:6}(b) and (d)].
In addition, we also mark the time scale associated with exchange field,
$t=1/|\exch |$, with vertical lines.

It can be seen in the time evolution of magnetization that,
independently of the coupling strength,
a minimum occurs at times $1\lesssim t\cdot U\lesssim 10$ or $10^{-1}\lesssim t\cdot \Gamma\lesssim1$,
where the magnetization points in the opposite direction
compared to its long-time-limit value.
Subsequently, a strong growth of magnetization is present
and the saturation is reached around $t\cdot U\gtrsim 10^2$ or $t\cdot\Gamma\gtrsim 10$.
The comparison of this behavior between panels (a) and (b) indicates
that the buildup of magnetization to good approximation
scales linearly with $\Gamma$.
Moreover, the saturation of magnetization also exhibits
the dynamics strongly dependent on $\Gamma$,
and for most considered values of the coupling, the maximum magnetization
is achieved at times $10\lesssim t\cdot \Gamma \lesssim 10^2$, see
Figs. \ref{Fig:6}(a) and (b).

Let us now discuss the time evolution of the evaluated exchange field $\exchn(t)$.
First of all, one can seen that the sign of the exchange field
is opposite to that of the induced magnetization, i.e.
we find $\exchn >0$ for $10^{-1}\lesssim t\cdot U \lesssim 10$
and $\exchn <0$ for $t\cdot U \gtrsim 10$, see \fig{Fig:6}(c).
Furthermore, as in the case of magnetization decreasing
the coupling strength generally results in larger values of $S_z(t)$,
in has just opposite effect on the generated exchange field.
It can be seen that the maximum value of $|\exchn(t)|$
decreases with lowering $\Gamma$.
This is in fact quite intuitive---the larger becomes the coupling to the ferromagnetic contact,
the larger the generated exchange field is.
Note, however, that for weaker couplings a relatively
low exchange field is sufficient to induce large magnetization in the quantum dot,
see \fig{Fig:6}(c).

To identify the relevant time scales for the sign change and the buildup of the exchange field,
in \fig{Fig:6}(d) we show $\exchn(t)/\Gamma$ plotted as a function of $t\cdot \Gamma$.
As can be seen in the inset, which presents the close-up
of $\exchn(t)$ where the sign change occurs, $\exchn(t)\approx 0$
for $t\cdot \Gamma \approx 1$, i.e. the sign change of the exchange field
develops for times of the order of $t\approx 1/\Gamma$. On the other hand,
it can be clearly seen that the time at which $|\exchn(t)|$
reaches its maximum does not scale linearly with $\Gamma$.
To estimate what is the scaling, we determine
the time $t_{\rm hmax}$ at which the absolute value of exchange field reaches
a half of its maximum value, $|\exchn(t_{\rm hmax})| \equiv {\rm max}\{|\exchn(t)|\}/2$.
In the inset of \fig{Fig:6}(c) we present
both $t_{\rm hmax}\cdot \Gamma$
and $t_{\rm hmax}\cdot \Gamma^2$ as a function of the coupling strength.
As results from these curves, the time associated
with the development of the exchange field scales
rather as $t_{\rm hmax} \propto \Gamma^2$
and not as $t_{\rm hmax} \propto \Gamma$.
This result is in fact quite counterintuitive, since from \eq{Eq:exch}
one could expect linear scaling of $\exchn(t)$ with the coupling strength.

\subsubsection{Influence of spin polarization}

\begin{figure}[t]
  \includegraphics[width=1\columnwidth]{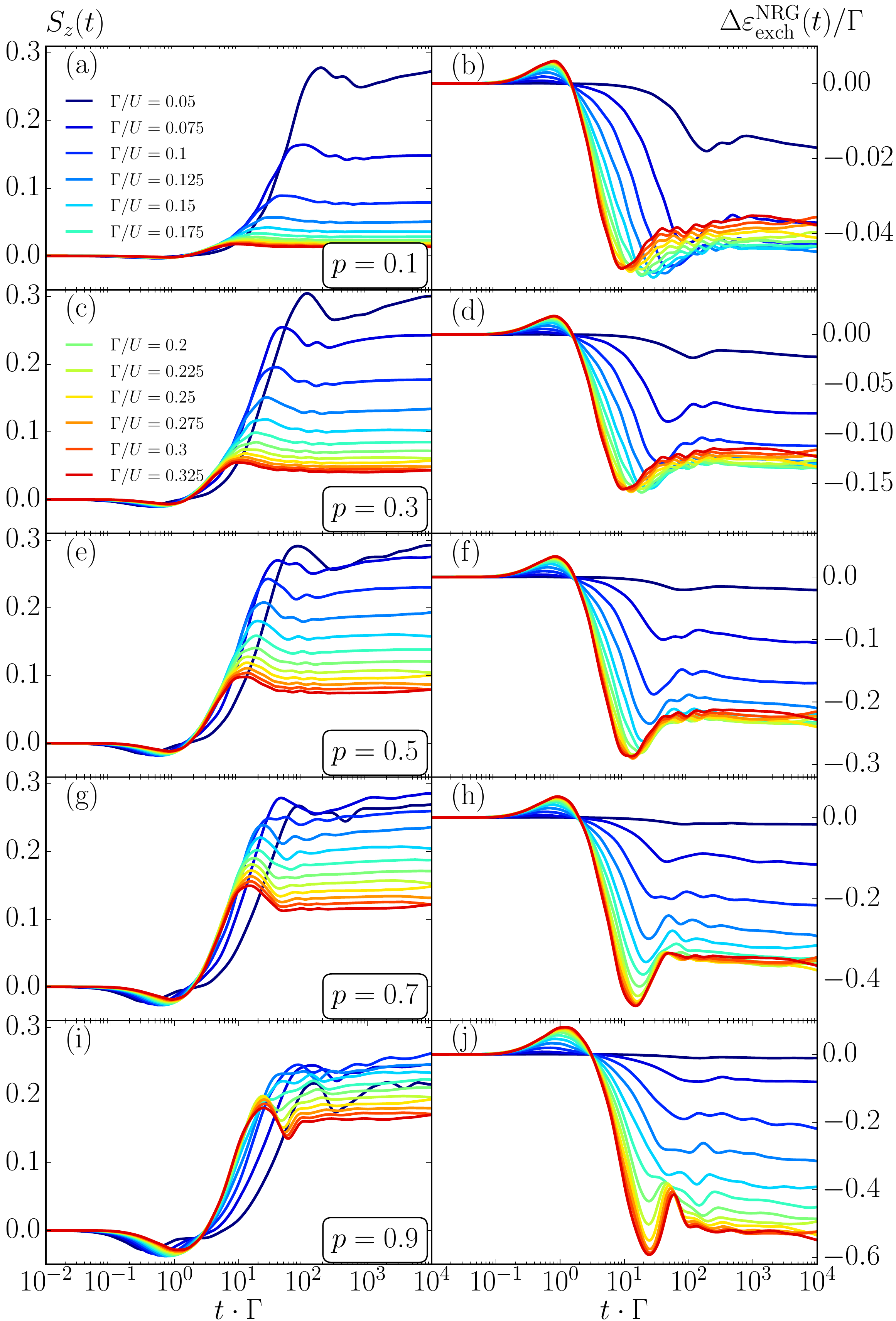}
  \caption{\label{Fig:7}
  (left column) The time-dependent magnetization $S_z(t)$
  and (right column) the generated exchange field $\exchn(t)$
  after the quench from an isolated quantum dot
  to the coupled regime calculated for
  different values of coupling strength $\Gamma$
  and the lead's spin polarization $p$, as indicated.
  The parameters are the same as in \fig{Fig:6}.
}
\end{figure}

The influence of the spin polarization $p$ of the ferromagnetic contact
on the spin dynamics is also nontrivial.
Figure \ref{Fig:7} presents the time evolution of the magnetization
(left column) and the exchange field (right column)
for different values of $p$ and $\Gamma$.
For relatively small values of spin-polarization, i.e. $p\lesssim 0.3$
[see panels (a) and (b) in \fig{Fig:7}],
neither magnetization nor exchange field
exhibit the sign change as a function of time.
This effect emerges once the spin polarization
becomes considerable, see the curves for $p\gtrsim 0.5$ in \fig{Fig:7}.
Moreover, with increasing $p$, the values of $S_z(t)$ and $\exchn(t)$
opposite to their long-time limits are increased.
Interestingly, the highest value of magnetization is obtained
for rather small values of $\Gamma$, almost independently of the spin polarization $p$.
Larger values of the coupling strength result in a faster dynamics
(the saturation occurs at earlier times), but on the other hand,
the long-time-limit value of magnetization gets lowered.
With increasing $p$, it is evident that the long-time limit
of magnetization and exchange field is enhanced, even for strong couplings, see \fig{Fig:7}.
Furthermore, one can clearly see that the magnitude of the exchange
field becomes enhanced with increasing the spin polarization.
In addition, for large values of $p$ the exchange field
$\exchn(t)$ depends more on the value of the coupling $\Gamma$,
cf. Figs. \ref{Fig:7}(b) and (j).

\subsubsection{Quantum dot's occupations}

\begin{figure}[t]
  \includegraphics[width=0.9\columnwidth]{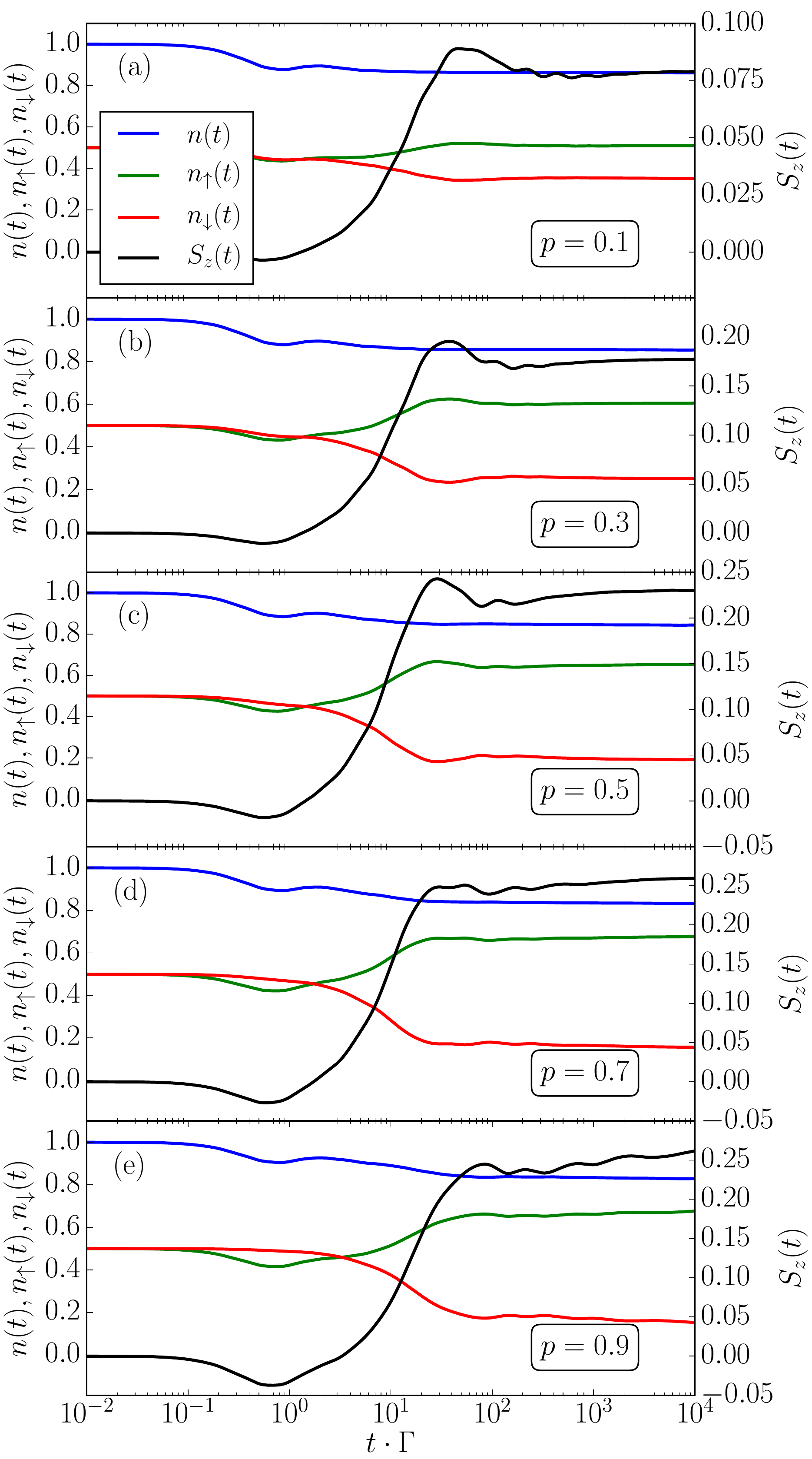}
  \caption{\label{Fig:10}
  The time-dependent expectation value of the local operators,
  $n(t)$, $n_\sigma(t)$ and $S_z(t)$,
  after the quench in the coupling strength
  calculated for selected values of the spin polarization of ferromagnetic contact $p$.
  The other parameters are the same as in \fig{Fig:6} with $\Gamma=U/10$.}
\end{figure}

Because one of the most interesting effects discussed here is the sign change of magnetization
and the associated exchange field, let us now focus on discussing the mechanism
responsible for this effect.
It turns out that the analysis of the expectation values of the corresponding
occupation operators $n(t)$, $n_{\uparrow}(t)$ and $n_{\downarrow}(t)$
can provide more detailed information about the spin dynamics of the system.
Figure \ref{Fig:10} presents the dot's occupations
$n(t)$, $n_{\uparrow}(t)$ and $n_{\downarrow}(t)$
calculated for different values of the ferromagnetic contact's spin polarization.
For comparison, we also show the time evolution of the dot's magnetization $S_z(t)$.
Clearly, increasing the spin polarization results in higher
values of magnetization in the long-time limit.
However, as already emphasized in the previous section,
the most interesting dynamics takes place at times around $t \approx 1/\Gamma $,
and it is generally associated with the difference
between the spin-resolved couplings $\Gamma^\uparrow$ and $\Gamma^\downarrow$
to the ferromagnetic contact.

First of all, one can see that the decrease of the total occupation $n(t)$
after the quench is similar, both qualitatively and quantitatively,
for all considered values of $p$. This decrease is the consequence
of the renormalization of the quantum dot level and its broadening
due to the coupling to external reservoir.
Note that in the figure $\e=-U/4$, such that $n(t>0)<1$.
It is thus clear that once the coupling is turned on, the total occupation
number of the dot becomes lowered as the time elapses
and it happens at short time scale, i.e. $n(t)$ starts
decreasing when $t\cdot \Gamma \approx 10^{-1}$
and for $t\cdot \Gamma \approx 1$, the total occupation
is already approximately equal to its long-time value, see \fig{Fig:10}.
It is however very important to consider how this precisely happens
as far as the spin-resolved occupations are concerned.
Because for finite $p$ the spin-up level is coupled
more strongly than the spin-down one, it is the spin-up
level that reacts first to the switching-on of the coupling.
Thus, at a short timescale, the occupation decrease
is mostly conditioned by the coupling $\Gamma^\uparrow$,
which leads to lowering of the occupation of the spin-up dot level.
However, eventually, the opposite spin component
with weaker coupling $\Gamma^\downarrow$ comes into play
and determines the dynamics of the system,
lowering its occupation accordingly,
as the magnetization grows and saturates for longer times.
This can be clearly seen in \fig{Fig:10}, especially for larger spin
polarizations---the drop of the total occupation
is mainly due to the decrease of $n_\uparrow(t)$,
such that one observes $n_\uparrow(t) < n_\downarrow(t)$ in a certain range of time.
However, as the time goes by,
the spin-dependence of charge fluctuations finally results
in equilibration, such that $n_\uparrow(t) > n_\downarrow(t)$.

In other words, the charge dynamics of the system is governed by the
stronger coupling to the majority-spin subband $\Gamma^\uparrow$, whereas
the spin dynamics is determined by the weaker coupling to the minority-spin subband
$\Gamma^\downarrow$.
Consequently, one observes a sign change of the magnetization
(and the induced exchange field) with the time range of magnetization
opposite to its long-time-limit value increased with enhancing the spin polarization $p$,
see \fig{Fig:10}.

\begin{figure}[t]
  \includegraphics[width=0.9\columnwidth]{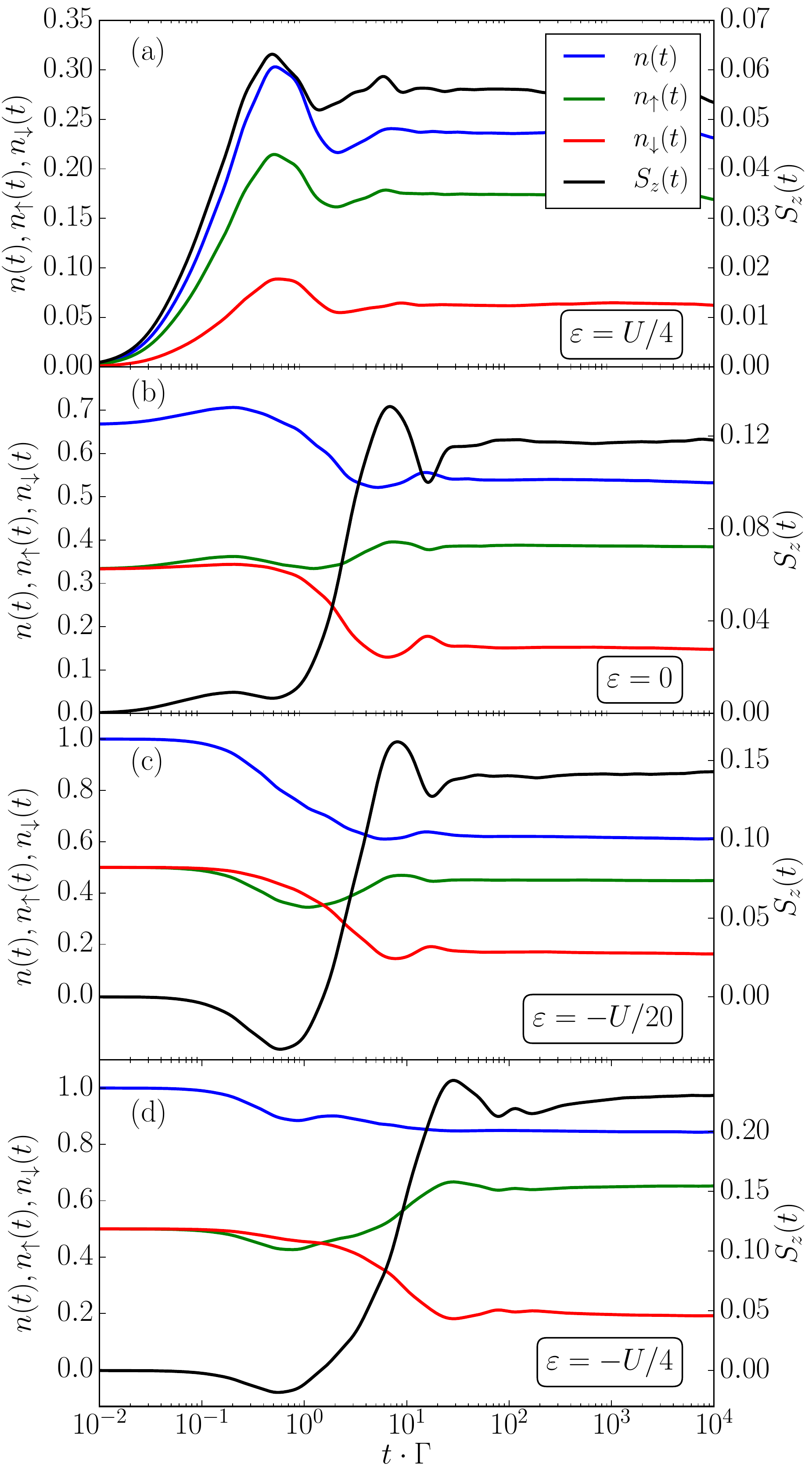}
  \caption{\label{Fig:9}
  The time-dependent expectation value of local operators
  after the quench in the coupling strength
  calculated for different values of the dots' energy level $\varepsilon$,
  as indicated, and for other parameters the same as in \fig{Fig:6}
  with $\Gamma=U/10$.
}
\end{figure}

In Fig. \ref{Fig:9} we show the relevant time evolution of the local operators
after the quench performed in the coupling strength
calculated for four different positions of the dot's energy level.
When in the initial state the quantum dot is empty, see
the case of $\e=U/4$ in \fig{Fig:9}(a),
the total occupation grows from $n(t=0)=0$ to around $n(t\to\infty)\approx0.25$ in the long-time limit.
Finite occupation after the quench is possible
due to the renormalization and broadening of the dot's energy level.
Due to the spin-dependent coupling, the occupation of the spin-up component
is higher with respect to the spin-down one, i.e. $n_{\uparrow}(t)>n_{\downarrow}(t)$,
which holds for all times $t>0$.
In consequence, the magnetization acquires only positive values
$S_z(t)>0$ and does not change sign at any positive time.
A similar behavior is in fact observed for $\e\geq 0 $.
For $\varepsilon=0$ [see \fig{Fig:9}(b)],
the initial occupation is non-zero, i.e. $n(0)=2/3$.
Then, switching on the coupling to the lead
results in the renormalization that decreases
the average occupation number, such that $n(t\rightarrow \infty) \approx 1/2$.
Note, however, that when the coupling is turned on
the total occupation first starts slightly increasing and then
decreases to reach one half. The behavior of
$n(t)$ is reflected in the dependence of the spin-resolved occupations.
The occupation of $n_\uparrow(t)$
exhibits small fluctuations as a function of time,
but in the long-time limit acquires a value
relatively close to the initial one, $n_\uparrow(t\rightarrow\infty)\approx 0.4$.
On the other hand, the evolution of $n_\downarrow(t)$ is strongly
correlated with the total occupation $n(t)$.
As a result, in this transport regime, the dot's magnetization is always positive.

However, when the energy of the orbital level is lowered further
such that in the initial state the dot is occupied by a single electron,
the spin dynamics gets qualitatively new features,
see Figs. \ref{Fig:9}(c) and (d).
As already explained earlier, now
the important effect of the renormalization and broadening due to
switching-on of the coupling is that the average occupation
of the quantum dot is decreased [$n(t) <1$] with respect to the initial state.
Moreover, the evolution of the system is now governed by two time scales,
while the first one, $t\sim 1/\Gamma^\uparrow$, is responsible for
charge dynamics, the second one, $t\sim 1/\Gamma^\downarrow$,
determines the magnetization dynamics.
The interplay between the two spin-resolved components of occupation results
in the oscillations of magnetization as a function of time with the corresponding sign change.
We also note that the strongest regime of opposite magnetization
occurs for $\varepsilon$ right below the Fermi level
and becomes lowered with further detuning the dot level towards
the particle-hole symmetry point, cf. Figs.~\ref{Fig:9}(c) and (d).

\subsection{Quench in the orbital level position}

\begin{figure}[t]
  \includegraphics[width=.95\columnwidth]{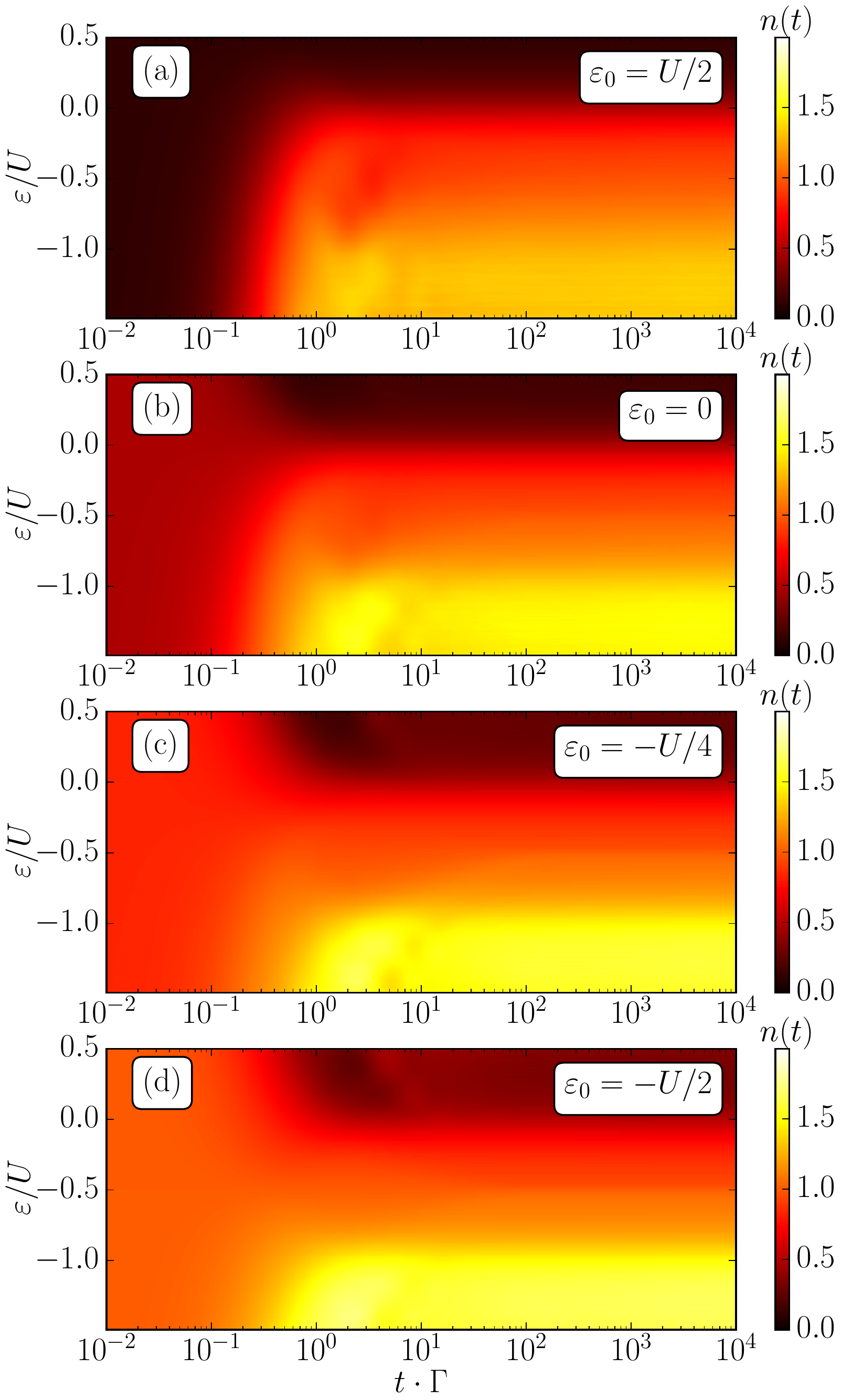}
  \caption{\label{Fig:11}
  The time-dependent occupation number $n(t)$
  after performing the quench in the quantum dot's orbital level position
  from $\varepsilon_{0}$ to $\varepsilon$.
  The other parameters are the same as in Fig.~\ref{Fig:6} with $\Gamma=U/10$.
}
\end{figure}

\begin{figure}[t]
  \includegraphics[width=.97\columnwidth]{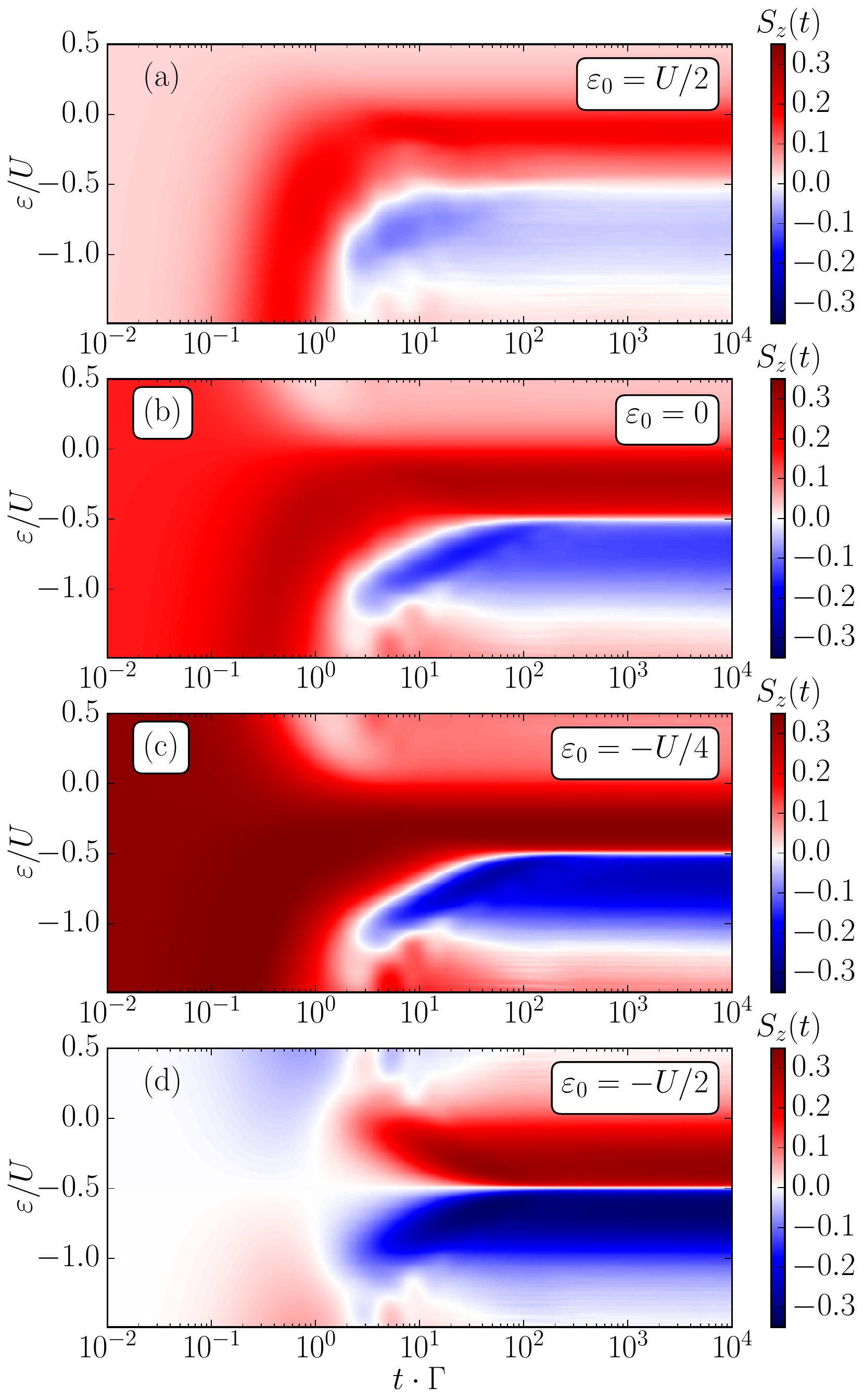}
  \caption{\label{Fig:12}
  The time-dependent magnetization $S_z(t)$ of the dot
  after performing the quench in the orbital level from $\varepsilon_{0}$ to $\varepsilon$.
  The other parameters are the same as in Fig.~\ref{Fig:6} with $\Gamma=U/10$.
}
\end{figure}

In this section we consider the quench performed in the position
of the quantum dot orbital level. The dot is coupled to the ferromagnetic lead
before the quench and the coupling strength remains unchanged, i.e. $\Gamma_0=\Gamma$.
The parameter that is abruptly switched at time $t=0$
is the dot's energy level $\varepsilon_{0} \rightarrow \varepsilon$.
We study the time evolution of the dot's occupation number (\fig{Fig:11})
and magnetization (\fig{Fig:12}) after the corresponding quench.
We consider four different initial energy levels $\varepsilon_{0}$
and the corresponding expectation values are calculated for a wide range
of final level position $\varepsilon$. Here, it is important to note
that the value of $\varepsilon_{0}$ determines
the quantum dot initial occupation number and magnetization.

As can be seen in \fig{Fig:11}, the short time evolution of the occupancy is mainly dependent on the initial occupation.
In all the considered cases, the occupation monotonically approaches
saturation at times $t\approx 1/\Gamma$.
Further behavior for times $t\gtrsim 1/\Gamma $ is qualitatively
similar across all values of $\varepsilon_{0}$ considered
and for all final level positions $\varepsilon$ approaches
the long-time limit. We note that there might occur a small deviation
of the long-time-limit value from the thermodynamical value,
which depends on the difference in energy between the initial and final Hamiltonians.
This is associated with the fact that, the larger this difference is,
it is more difficult for the system to dissipate energy in the long-time limit,
which is a direct consequence of the fact that the system
does not fully thermalize on the finite Wilson chain \cite{Rosch2012,Weymann2015}.
When the quench has a relatively large energy difference,
i.e. $|\varepsilon_{0} - \varepsilon| \gtrsim U$, an oscillatory behavior
is visible right after attaining the maximum value
at times $1 \lesssim t\cdot \Gamma \lesssim 10$,
for $\varepsilon/U \lesssim -1$ and $\varepsilon/U \gtrsim 0.5$,
see Figs. \ref{Fig:11}(c) and (d).

The quench dynamics is even more interesting when the time
evolution of quantum dot's magnetization $S_z(t)$ is considered.
Now, the initial position of the dot's energy level $\varepsilon_0$ strongly determines
the behavior of the magnetization for short times ($t\cdot \Gamma\lesssim 10^{-1}$).
In general, independently of the initial conditions,
for the final values of the energy level above the particle-hole symmetry point,
i.e. $\varepsilon>-U/2$, the quantum dot acquires
magnetization, which is parallel to the magnetization of the ferromagnet.
For the particle-hole symmetry point ($\varepsilon=-U/2$),
the exchange field vanishes and the magnetization does not develop.
On the other hand, for the dot level position below the particle-hole symmetry point,
$\varepsilon<-U/2$, the exchange field changes sign and
the quantum dot is magnetized in the opposite direction.

Let us now discuss the system's dynamics in more detail and focus
on the influence of the initial condition, i.e. the value of $\varepsilon_0$,
on the time dependence of the dot's occupation and magnetization.
For the orbital level set above the Fermi level, see \fig{Fig:11}(a) where $\varepsilon_0=U/2$,
the initial occupation of the quantum dot is small but finite
$n(0)\approx 0.1$, which in consequence results in a finite magnetization
in the direction of the magnetization of the ferromagnetic lead, see \fig{Fig:12}(a).
At times $t\cdot \Gamma \gtrsim 10^{-1}$, the magnetization starts to growth.
Further time dependence of $S_z$ significantly depends on
the final level position $\varepsilon$.
For $\varepsilon>0$, the quantum dot
mildly and monotonically increases its occupation number and, accordingly,
the magnetization grows in a similar manner.
However, when $0>\varepsilon>-U/2$, the magnetization buildup is rapid
compared to the previous regime,
which is due to higher occupation number $n(t)\approx1$.
Here, the dynamics of charge and spin are very similar
as both average expectation values saturate at times $t\approx 1/ \Gamma $.
On the other hand, for $\varepsilon<-U/2$,
the system magnetizes in the opposite direction.
Now, when the initial position of the dot level is shifted toward lower energies,
see Figs.~\ref{Fig:11}(b)-(c) and \ref{Fig:12}(b)-(c),
one can observe two effects.
Firstly, the initial magnetization is stronger as $\varepsilon_0$ is lowered,
which is due to an increased occupation at the initial state.
This is visible down to the particle-hole symmetry point,
cf. Figs. \ref{Fig:11}(d) and \ref{Fig:12}(d).
Secondly, the long-time-limit magnetization is strongly enhanced.
When lowering the initial position of the orbital level further,
the quench is performed from the lower-energy state and therefore,
it is easier for the system to achieve the thermal average in the long-time limit.

Finally, we consider the case when the dot is set at the particle-hole symmetry point in the initial state,
where $S_z(t=0)=0$. In general, in this case the spin dynamics is antisymmetric
with respect to detuning from the particle-hole symmetry point,
see \fig{Fig:12}(d).
The quantum dot is initially spin unpolarized $S_z(0)=0$
and for a wide range of the final position of the orbital level energy $\varepsilon$
it starts to build up magnetization for times $10^{-1} \lesssim t\cdot\Gamma \lesssim 1$
in the opposite direction to its long-time-limit value.
Consequently, for times $1 \lesssim t\cdot\Gamma \lesssim 10$,
there is at least one sign change present in the case of $0<\varepsilon/U<-1$
(except for the particle-hole symmetry point and its vicinity).
Moreover, an oscillatory behavior of the magnetization
takes place in the case of stronger quenches,
i.e for $\varepsilon>0$ or $\varepsilon<-U$.
In the above regimes of $\varepsilon$, the absolute value of the long-time limit
of magnetization is also lower compared to the magnetization in the singly occupied regime,
see \fig{Fig:12}(d).
The long-time-limit value of the dot's magnetization is fully suppressed for the transport regime
with doubly occupied and empty quantum dot $S_z(t\rightarrow \infty)=0$,
which is visible for $\varepsilon=U/2$ and $\varepsilon=-3U/2$ in \fig{Fig:12}(d).

\subsection{Finite temperature effects}

\begin{figure}[t]
  \includegraphics[width=0.9\columnwidth]{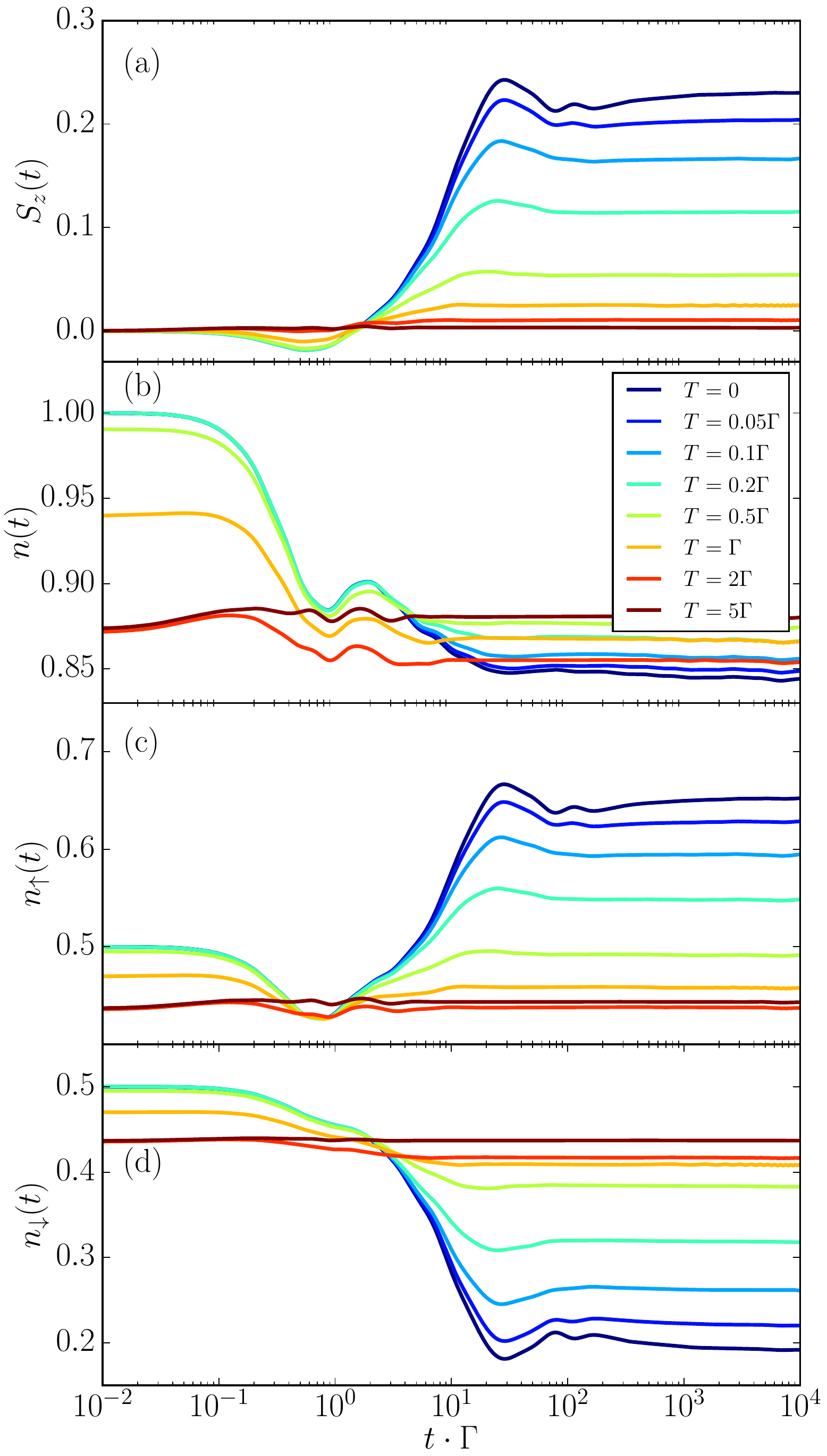}
  \caption{\label{Fig:13}
  The expectation values of local operators after the quench
  in the coupling strength from $\Gamma_0 = 0$
  to $\Gamma = U/10$ plotted as a function of time
  and calculated for different temperatures, as indicated.
  The other parameters are the same as in \fig{Fig:6}.
}
\end{figure}

\begin{figure}[t]
  \includegraphics[width=0.9\columnwidth]{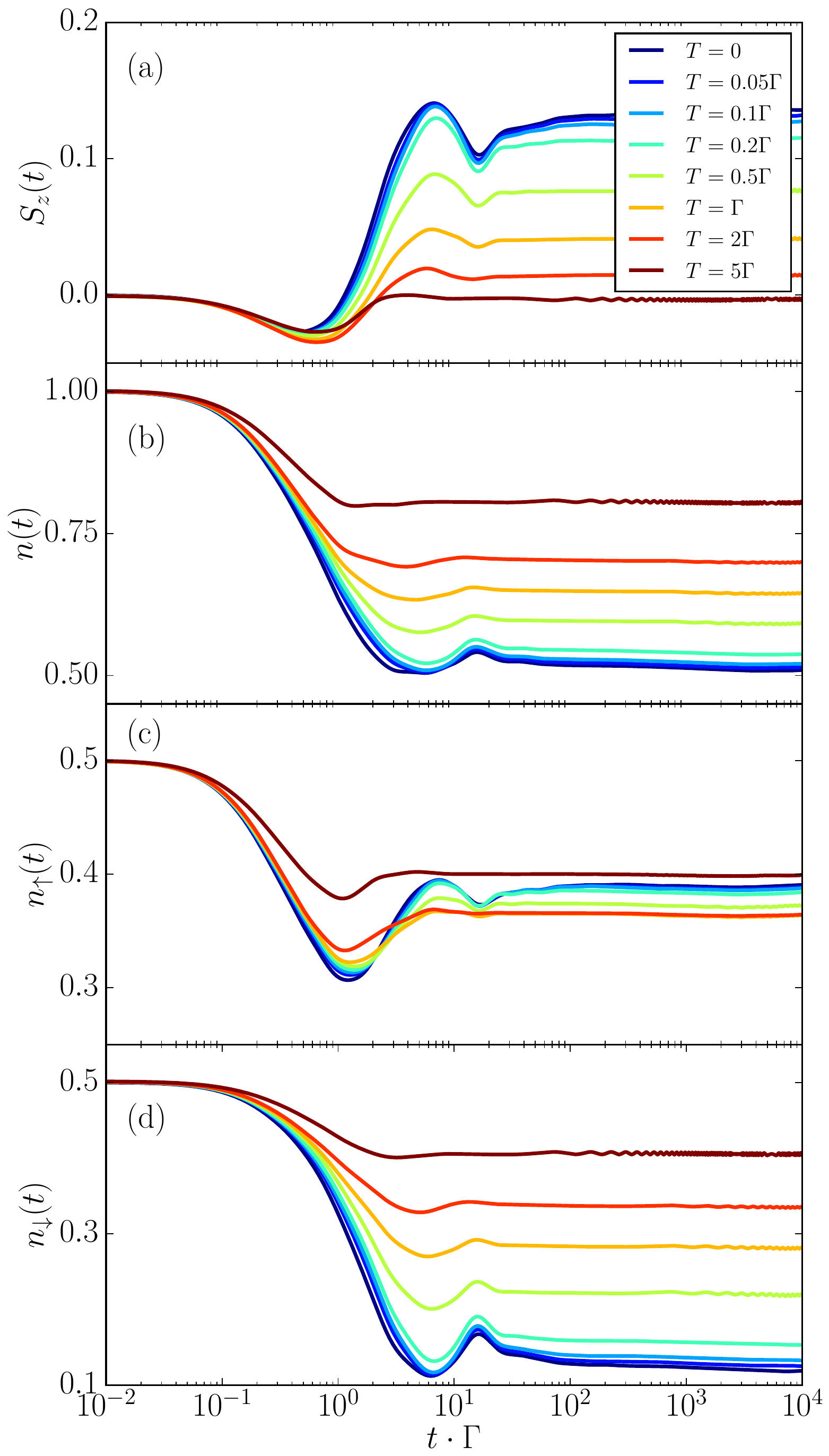}
  \caption{\label{Fig:14}
  The same as in \fig{Fig:13} calculated
  for the quench in the quantum dot occupation
  from $\varepsilon_0=-U/2$ to $\e=0$.
  The other parameters are the same as in \fig{Fig:6}
  with $\Gamma = U/10$.}
\end{figure}

Let us now consider the influence of finite temperature $T$ on
the dynamics of the system, which undergoes quenches discussed in the preceding sections.
We focus on the most interesting case with a single
electron occupying the quantum dot.
Figure~\ref{Fig:13} presents the time evolution of local operators
after the quench in the coupling strength calculated
for different values of temperature $T$ expressed in the units of $\Gamma$ ($k_B\equiv 1$).
It can be seen that at zero temperature the dot occupation
slightly decreases due to the fact that the system is detuned from the particle-hole symmetry point
($\e=-U/4$ in the figure).
The different time-dependence of the spin-resolved occupations
results in finite magnetization, which changes sign
around $t \approx 1/\Gamma$, as explained in the previous sections.
When the temperature is increased, the long-time value of the magnetization
becomes strongly suppressed and for temperatures of the order of the coupling strength,
$S_z(t)\approx 0$. This is associated with the fact that
the spin-resolved charge fluctuations become overwhelmed by thermal fluctuations,
which essentially suppresses the system dynamics once $t\gtrsim 1/T$.
More specifically, with increasing the temperature,
the difference in the total occupation between the initial and final states strongly drops,
see \fig{Fig:13}(b). For $T=0$, the quench modifies the occupation number
from $n(t=0)=1$ to $n(t\cdot \Gamma>10)\approx 0.85$,
while for finite temperatures the difference between the initial
and long-time-limit value of the occupation is much smaller due to enhanced thermal fluctuations.
Moreover, thermal fluctuations are responsible for decreasing
the difference in the occupation of the spin-up and spin-down components,
which is clearly visible when one compares panels (c) and (d) in \fig{Fig:13}.
This altogether leads to the suppression
of the dot's magnetization and, consequently, the induced exchange field.

The case when the quench is performed in the dot's orbital level
is presented in Fig.~\ref{Fig:14}. We consider the scenario when initially
the system tuned to the particle-hole symmetry point.
Therefore, the initial magnetization is equal to $S_z(t=0)=0$,
while the occupation number is given by $n(t=0)=1$.
Then, the orbital level is detuned from this point to $\e=0$,
such that finite magnetization builds up in the dot as the times elapses.
At first, the dependence is qualitatively very similar to the previous case,
where the coupling strength was quenched,
cf. Figs.~\ref{Fig:13}(a) and \ref{Fig:14}(a).
The long-time limit of magnetization drops as temperature is
increased in a similar fashion. However, there is a qualitative difference,
since now a higher temperature is necessary to fully suppress the magnetization.
This is related to the energy difference between
the initial and final Hamiltonians describing the quench,
which in the case of quench in the orbital level position
is larger than in the previous quench by around one order of magnitude.
The influence of finite temperature is clearly visible in \fig{Fig:14}(b),
where the long-time-limit value of the occupation is enhanced with $T$.
As far as the spin-dependent components are concerned,
the effect of thermal fluctuations is relatively weak on the spin-up occupation,
while it mainly increases the occupation of the spin-down occupation,
see Figs.~\ref{Fig:14}(c) and (d).
Altogether, finite temperature balances both spin-resolved
components of the dot's occupation and leads
in consequence to the drop of the dot's magnetization, see \fig{Fig:14}(a).


\section{Conclusions} \label{conclusions}

In this paper we have examined the spin-resolved quench dynamics
of a correlated quantum dot attached to a reservoir of spin-polarized electrons.
The considerations were performed by using the
time-dependent numerical renormalization group method
in the matrix product state framework. We studied
the system dynamics by considering two types of quantum quenches:
the first one was performed in the coupling strength,
whereas the second one was performed in the position of the dot's orbital level.
The emphasis was put on the analysis of the time-dependent
expectation values of local operators, such as the dot's occupation number and magnetization.
By comparing the induced magnetization with
the expectation value of the dot's spin for nonmagnetic contacts
in the presence of magnetic field, we were able to estimate
the magnitude of generated exchange field and analyze its buildup in time.
Moreover, by implementing the full density matrix of the system, we
have also examined the effects of finite temperature on the spin dynamics.

In the case of quench performed in the coupling strength we carried out
a detailed analysis of the influence of the quantum dot initial occupation
on the time evolution of the dot's magnetization and occupation.
In particular, we found a time range where a sign change
occurs during the nonmonotonic build up of magnetization
and the associated induced exchange field.
We identified two time scales describing this
nontrivial spin dynamics, and explained this effect
by performing a detailed analysis of the time-dependence
of expectation values of spin-resolved quantum dot occupations.
It turned out that while the charge dynamics is mainly governed
by the coupling to majority spin subband of the ferromagnet,
the spin dynamics is mostly determined by the minority-spin-subband coupling.
This results in qualitatively different time-dependence
of spin-resolved quantum dot's occupations,
which reveals through the corresponding sign change of the magnetization.

Furthermore, the case of quench performed in the dot's orbital level position was considered.
Similarly to the first type of quench,
we accentuated the influence of the system's initial conditions
on the system's dynamical behavior.
Despite relatively clear and simple time dependence of the quantum dot total occupancy,
we found the spin dynamics to be nontrivial.
In particular, we showed that the system quenched from the particle-hole symmetry point
exhibits a nonmonotonic behavior of magnetization that can include multiple sign changes.

In addition, we have analyzed the influence of finite
temperature on both types of the considered quenches.
The thermal fluctuations strongly
suppress the dynamics of the system for times $t\gtrsim 1/T$.
More specifically, finite temperature is responsible for balancing the spin-up
and spin-down components of the quantum dot occupation,
which is clearly visible as a drop of the dot's magnetization.

Finally, we note that while the exchange field in the long-time limit
can be seen as an effective magnetic field acting on the dot \cite{Gaass2011},
at shorter times, of the order of $t\approx 1/\Gamma$,
it results in an interesting dynamical behavior of the system
involving a sign change of the quantum dot magnetization.
In this case intuitive analogy to simple application
of external magnetic field is rather unjustified.


\begin{acknowledgments}
We gratefully acknowledge discussions with Andreas Weichselbaum.
This work was supported by the Polish National Science
Centre from funds awarded through the decision No. 2017/27/B/ST3/00621.
Computing time at the Pozna\'n Supercomputing
and Networking Center is also acknowledged.
\end{acknowledgments}


\bibliography{Bibliography}{}

\end{document}